\documentclass[12pt]{article}

\usepackage{amsmath,amssymb}
\usepackage{graphicx}
\usepackage{amscd}
\usepackage{theorem}

\newcommand{\R}{{\mathbb{R}}}
\newcommand{\Z}{{\mathbb{Z}}}
\newcommand{\C}{{\mathbb{C}}}
\newcommand{\T}{{\mathbb{T}}}

\newcommand{\ba}{\begin{array}}
\newcommand{\ea}{\end{array}}

\newcommand{\bp}{\begin{pmatrix}}
\newcommand{\ep}{\end{pmatrix}}

\newcommand{\bps}{\begin{smallmatrix}}
\newcommand{\eps}{\end{smallmatrix}}

\newcommand{\s}{\sqrt}

\def \({\left(}
\def \){\right)}

\def \cA{{\cal A}}

\DeclareMathOperator{\Tr}{Tr}

\def \raw{\rightarrow}

\def \Im{\mathrm{Im}}

\def \Ed#1#2{{\mbox End}_{#1} {#2}}

\def \ov#1{\frac{1}{#1}}

\def \cA{{\cal A}}

\def \Eh{\hat{E}}

\def \fpartial#1{\frac{\partial}{\partial {#1}}}

\def \l{{\frak l}}

\def \l({\left(}
\def \r){\right)}

\def \0{{\bf 0}}
\def \1{{\bf 1}} 

\def \Mat{\mathit{Mat}}
\def \ie{{\it i.e.\ }}

\begin{document}

\begin{titlepage}
\thispagestyle{empty}
\begin{flushleft}
\hfill hep-th/0207097\\
\hfill July, 2002 \\
\end{flushleft}

\vskip 1.5 cm

\begin{center}
\noindent{\Large \textbf{Kronecker foliation, D1-branes and}}\\

\vspace*{0.7cm}

\noindent{\Large \textbf{Morita equivalence of Noncommutative two-tori}}\\

\noindent{
 }\\
\renewcommand{\thefootnote}{\fnsymbol{footnote}}

\vskip 2cm
{\large 
Hiroshige Kajiura 
\footnote{e-mail address: kuzzy@ms.u-tokyo.ac.jp}\\

\noindent{ \bigskip }\\

\it
Graduate School of Mathematical Sciences, University of Tokyo \\
3-8-1 Komaba, Tokyo, 153-8914 Japan\\
\noindent{\smallskip  }\\
}

\bigskip
\end{center}
\begin{abstract}

It is known that the physics of open strings on a D2-brane on a two-torus 
is realized from the viewpoint of deformation quantization 
in the Seiberg-Witten limit. 
We study its T-dual theory, \ie D1-brane physics on two-tori. 
Such theory is described by Kronecker foliation. 
The algebra of open strings on the D1-brane is then identified 
with the crossed product representation of a noncommutative two-torus. 
The Morita equivalence of noncommutative two-tori is also 
realized geometrically along this line. 
As an application, 
Heisenberg modules and the tensor product between them are 
discussed from these geometric viewpoints. 
We show they are related to the homological mirror symmetry of two-tori. 

\end{abstract}
\vfill

\end{titlepage}
\vfill
\setcounter{footnote}{0}
\renewcommand{\thefootnote}{\arabic{footnote}}
\newpage

\tableofcontents

\section{Introduction}

Noncommutative geometry is directly related to 
the physics of open strings \cite{CDS}-\cite{CF}. 
A noncommutative torus is an ideal example of noncommutative algebras. 
It is known that noncommutative tori have a beautiful symmetry, 
the Morita equivalence\cite{Rtwo}-\cite{S}. 
The Morita equivalence is an important concept also for physics. 
It is the equivalence of (noncommutative) algebras 
which have the isomorphic category of projective modules
\footnote{There exists another but equivalent (more explicit) definition 
of the Morita equivalence, which is stated in Subsection \ref{ssec:pm}. 
}. 
Here, in the spirit of the noncommutative geometry, 
a noncommutative algebra is regarded as the space of functions on a 
`noncommutative space'.
The projective modules over a noncommutative algebra are then 
vector bundles, \ie D-branes 
over the noncommutative space. 
Therefore one can say physically that the Morita equivalent algebras 
are the noncommutative target spaces on which the same set of D-branes 
exists. Furthermore, 
if one has a background independent field theory of open strings, 
each D-brane should correspond to each classical solution of the 
field theory. Actually, for the effective field theory of open strings 
on (two-)tori, it is shown that each classical solution corresponds to 
each brane\cite{BKMT,KMT} in the context of tachyon condensation
\cite{tac}. 
Thus, the investigation of noncommutative algebras and their Morita 
equivalence is an important subject which is relevant to 
the nonperturbative effects and furthermore the background independence 
of the field theory of open strings. 
In addition, noncommutative expressions have the advantage that 
projective modules over a noncommutative space 
can uniformly describe D-branes 
including those which cannot be described by vector bundles. 
For instance, 
D0-brane on a two-torus with no D2-brane cannot be expressed 
in terms of a vector bundle 
over the two-torus, but can be expressed in terms of 
a projective module over a noncommutative-torus. 

Noncommutative tori are represented in several ways; 
deformation quantization\cite{BFFLS}, 
irrational rotation algebra\cite{co-book,CR}, and so on. 
For simplicity, let us concentrate on two-dimensional tori. 
Open string theory on a D2-brane can be realized from the viewpoint of 
the deformation quantization\cite{Scho,SW}, 
so it is described by a noncommutative two-torus. 
Morita equivalent noncommutative two-tori are then generated 
by group $SL(2,\Z)$. 
It is known to be related to T-duality group, so 
the Morita equivalent noncommutative two-tori 
can be regarded as the algebras 
of gauge fields on D2-D0 brane bound states\cite{CDS,Ho,MZ,BMZ}. 
On the other hand, D1-brane physics on a two-torus is not 
Morita equivalent to the D2-brane physics. 
The D1-brane physics is, however, T-dual to the D2-brane (D2-D0) physics. 
(In the rest of this paper we use the term `T-duality' in this sense.) 
Geometrically, the D1-brane picture is simpler, so is 
used to realize the D2-brane physics and the $SL(2,\Z)$ symmetry 
which acts on the D2-brane physics\cite{AAS,I,K}. 
The noncommutativity of the algebra of noncommutative two-tori 
is also understood intuitively in this picture\cite{DH}. 
However, the noncommutativity in the D1-brane picture is not 
described by the deformation quantization (Moyal-product). 

In this paper, the algebra of open string field on the D1-brane is 
identified with the crossed product representation of noncommutative tori. 
Namely, a noncommutative torus represented in the deformation quantization 
picture and the one represented by crossed product are 
connected by T-duality. 
On the D1-brane physics, the crossed product is obtained from the geometry 
of the Kronecker foliation. 
The Morita equivalence of the algebra of open strings is 
then realized naturally in this picture. 
Note that for two-tori the T-dual is equivalent to mirror dual. 
Homological mirror symmetry\cite{mirror} is then the open string 
version of the mirror symmetry\cite{W2} from a physical viewpoint. 
Thus, the results stated above are expected to be 
relevant to the homological mirror symmetry on two-tori. 
In fact, from the viewpoints of the results, 
we succeed to find a natural subcategory of the 
category of projective modules 
over noncommutative two-tori which corresponds to a category in 
homological mirror symmetry setup. 
The fact might give an insight on the nonperturbative structure 
over noncommutative tori. 
The subcategory is closely related to the theory of 
holomorphic vectors (or theta vectors) 
introduced by A.~Schwarz\cite{Stheta,Stensor}.

This paper is organized as follows. 
In Section 2, we review the basic facts on the Kronecker foliation. 
Subsection 2.1 explains the relation between 
the Kronecker foliation on a two-torus and the crossed product 
representation of a noncommutative two-torus. 
The Kronecker foliation provides us an intuitive picture 
why Morita equivalent noncommutative two-tori are generated by $SL(2,\Z)$, 
which is explained in Subsection 2.2. 
In Section 3 we consider D1-branes on the two-torus and 
the physics of the open string ending on them. 
Subsection 3.1 shows that the algebra of the open strings on the D1-brane 
is just the noncommutative two-torus in the crossed product representation 
explained in Subsection 2.1. 
The D1-branes can wrap various cycles $S^1$ on the two-torus. 
We discuss the open strings on such D1-branes in Subsection 3.2. 
Such situations correspond to that 
in Subsection 2.2 in a certain limit. 
We discuss in Section 4 when the situations 
in Subsection 3.2 and Subsection 2.2 coincide. 
It is shown that both situations coincide in the limit related to 
the so-called Seiberg-Witten limit\cite{SW}. 
Section 5 presents an application of the results stated above. 
In Subsection 5.1 projective modules on noncommutative two-tori 
are realized along the previous arguments. 
In Subsection 5.2 we relate the theory on the projective modules to 
the homological mirror symmetry. 
Namely, noncommutative analogue of the homological mirror 
symmetry is proposed. Alternatively, 
the results also mean that some open string interaction in 
the noncommutative theory can be realized as disk instanton contributions.

 \section{Kronecker foliation and Morita equivalence}
\label{sec:2}

Noncommutative tori can be obtained from the Kronecker foliation
\cite{co-book}. 
We shall review the fact in Subsection \ref{ssec:21}. 
Such a representation 
of noncommutative tori 
admits an intuitive and geometric realization 
for the Morita equivalence. 
We shall explain it in Subsection \ref{ssec:22}.  

 \subsection{Kronecker foliation and noncommutative torus}
\label{ssec:21}

Let $(x_1,x_2)$ be the coordinates on a two-torus $\T^2\simeq \R^2/\Z^2$ 
with periodicity $x_1\sim x_1+1$ and $x_2\sim x_2+1$. 
Let us consider the line 
\begin{equation}
 x_2=\theta x_1
 \label{irrslope}
\end{equation}
on the covering space $\R^2$. 
When the slope $\theta$ is rational, the image of the line 
on $\T^2$ is $S^1$. 
However when $\theta$ is irrational, 
the image fills densely in $\T^2$. 

\begin{figure}[h]
 \hspace*{3.0cm}\includegraphics{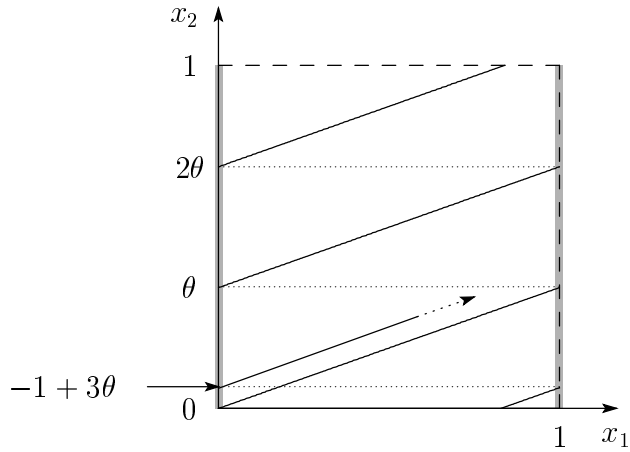}
 \caption{The Kronecker foliation with slope $\theta$. 
The line of slope $\theta$ is the leaf of the foliation. The leaf is 
transversal to the cycle $S^1$ described by $x_1=c_1, c_1\in\Z$. }
 \label{fig:f1}
\end{figure}

In this paper, we shall discuss the irrational case. 
The pair of $\T^2$ and the line (\ref{irrslope}) on $\T^2$ 
(Fig.\ref{fig:f1}) 
is called the Kronecker foliation with the irrational slope. 
Generally, a manifold $M$ is said to be foliated when we have 
a partition of $M$ into 
its submanifolds whose codimension is greater than one. 
Such submanifolds are then called the leaves of $M$. 
Here the line Eq.(\ref{irrslope}) is the leaf of the Kronecker foliation. 

Let us parametrize the leaf as $(x_1, x_2)=(t,\theta t)$, $t\in\R$. 
Now we consider $S^1$ defined by $x_1=0$ and functions 
on the $S^1$. The functions are represented 
as a Fourier-expanded form : 
\begin{equation*}
 a(x_2)=\sum_{m\in\Z}a_m e^{2\pi i x_2}\ .
\end{equation*}
The generator of these functions is $e^{2\pi i x_2}$ and 
we define 
\begin{equation*}
 U_2:=e^{2\pi i x_2}\ .
\end{equation*} 
On the covering space of $\T^2$, there are the mirror images of $S^1$ 
which are expressed as $x_1=c_1, c_1\in\Z$. 
Note that the leaf is necessarily transversal to the $S^1$ 
as in Fig.\ref{fig:f1}. 
The point $t=0$ is an intersection point of the leaf and the $S^1$, 
and the leaf then intersects with the $S^1$ in the next time at $t=1$, 
\ie $(x_1,x_2)=(1, \theta)$. 
Correspondingly, we define the following action $U_1$ 
on the functions on $S^1$. 
\begin{equation*}
 U_1 a(x_2)=a(x_2+\theta)\ .
\end{equation*}
These two generators satisfy the following relation
\begin{equation*}
 U_1 U_2=e^{2\pi i\theta}U_2 U_1\ .
\end{equation*}
Thus, the algebra generated by $U_1$ and $U_2$ is 
the noncommutative two-torus.
Such a representation of the noncommutative torus is called 
the crossed product representation of the rotation algebras\cite{co-book}.

 \subsection{Kronecker foliation and Morita equivalence}
\label{ssec:22}

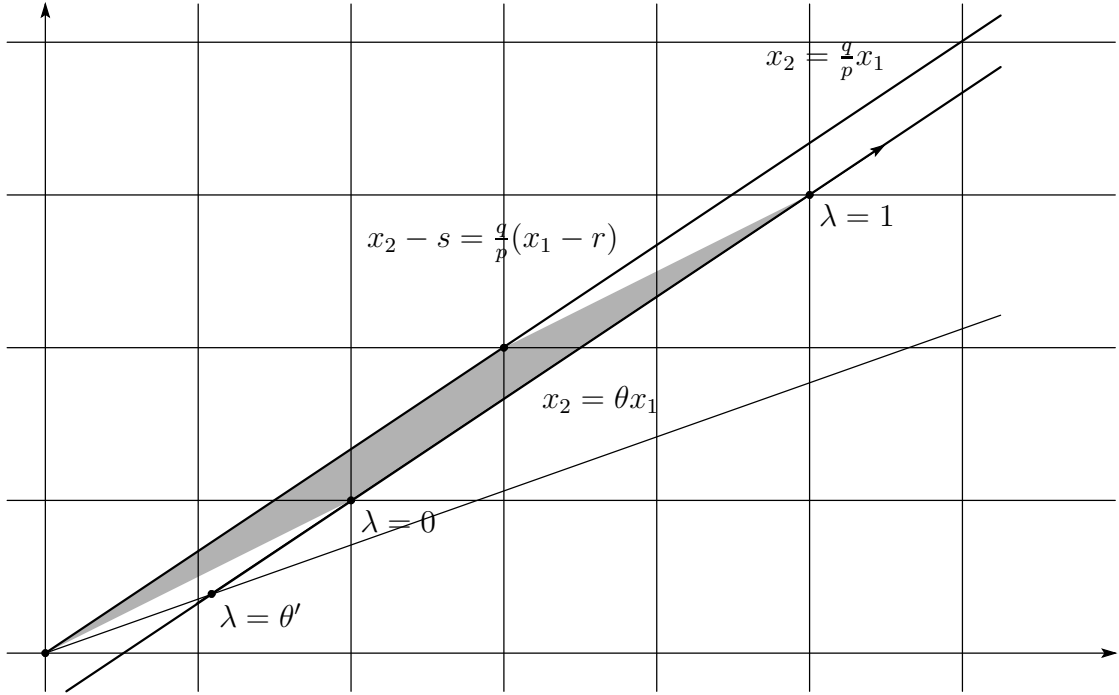
\begin{figure}[h]
\unitlength 0.1in
\begin{picture}( 58.0000, 36.0000)(  8.0000,-38.0000)
%
\special{pn 8}%
\special{sh 0.300}%
\special{pa 1000 3600}%
\special{pa 3400 2000}%
\special{pa 5000 1200}%
\special{pa 2600 2800}%
\special{pa 1000 3600}%
\special{pa 1000 3600}%
\special{pa 1000 3600}%
\special{ip}%
%
\special{pn 8}%
\special{pa 800 3600}%
\special{pa 6600 3600}%
\special{fp}%
\special{sh 1}%
\special{pa 6600 3600}%
\special{pa 6534 3580}%
\special{pa 6548 3600}%
\special{pa 6534 3620}%
\special{pa 6600 3600}%
\special{fp}%
%
\special{pn 8}%
\special{pa 1000 3800}%
\special{pa 1000 200}%
\special{fp}%
\special{sh 1}%
\special{pa 1000 200}%
\special{pa 980 268}%
\special{pa 1000 254}%
\special{pa 1020 268}%
\special{pa 1000 200}%
\special{fp}%
%
\special{pn 8}%
\special{pa 800 400}%
\special{pa 6600 400}%
\special{fp}%
%
\special{pn 8}%
\special{pa 800 1200}%
\special{pa 6600 1200}%
\special{fp}%
%
\special{pn 8}%
\special{pa 800 2000}%
\special{pa 6600 2000}%
\special{fp}%
%
\special{pn 8}%
\special{pa 800 2800}%
\special{pa 6600 2800}%
\special{fp}%
%
\special{pn 8}%
\special{pa 1800 3800}%
\special{pa 1800 200}%
\special{fp}%
%
\special{pn 8}%
\special{pa 2600 3800}%
\special{pa 2600 200}%
\special{fp}%
%
\special{pn 8}%
\special{pa 3400 3800}%
\special{pa 3400 200}%
\special{fp}%
%
\special{pn 8}%
\special{pa 4200 3800}%
\special{pa 4200 200}%
\special{fp}%
%
\special{pn 8}%
\special{pa 5000 3800}%
\special{pa 5000 200}%
\special{fp}%
%
\special{pn 8}%
\special{pa 5800 3800}%
\special{pa 5800 200}%
\special{fp}%
%
\special{pn 13}%
\special{pa 6000 530}%
\special{pa 1110 3800}%
\special{fp}%
%
\special{pn 13}%
\special{pa 1000 3600}%
\special{pa 6000 260}%
\special{fp}%
%
\special{pn 20}%
\special{sh 1}%
\special{ar 1870 3290 10 10 0  6.28318530717959E+0000}%
\special{sh 1}%
\special{ar 1870 3290 10 10 0  6.28318530717959E+0000}%
%
\special{pn 8}%
\special{pa 1790 3340}%
\special{pa 5390 940}%
\special{fp}%
\special{sh 1}%
\special{pa 5390 940}%
\special{pa 5324 960}%
\special{pa 5346 970}%
\special{pa 5346 994}%
\special{pa 5390 940}%
\special{fp}%
\put(50.5000,-12.5000){\makebox(0,0)[lt]{$\lambda=1$}}%
%
\special{pn 20}%
\special{sh 1}%
\special{ar 5000 1200 10 10 0  6.28318530717959E+0000}%
\special{sh 1}%
\special{ar 5000 1200 10 10 0  6.28318530717959E+0000}%
%
\special{pn 20}%
\special{sh 1}%
\special{ar 2600 2800 10 10 0  6.28318530717959E+0000}%
\special{sh 1}%
\special{ar 2600 2800 10 10 0  6.28318530717959E+0000}%
\put(26.5000,-28.5000){\makebox(0,0)[lt]{$\lambda=0$}}%
%
\special{pn 20}%
\special{sh 1}%
\special{ar 1000 3600 10 10 0  6.28318530717959E+0000}%
\special{sh 1}%
\special{ar 1000 3600 10 10 0  6.28318530717959E+0000}%
%
\special{pn 20}%
\special{sh 1}%
\special{ar 3400 2000 10 10 0  6.28318530717959E+0000}%
\special{sh 1}%
\special{ar 3400 2000 10 10 0  6.28318530717959E+0000}%
%
\special{pn 8}%
\special{pa 1000 3600}%
\special{pa 6000 1830}%
\special{fp}%
\put(19.1000,-33.4000){\makebox(0,0)[lt]{$\lambda=\theta'$}}%
\put(36.0000,-22.0000){\makebox(0,0)[lt]{$x_2=\theta x_1$}}%
\put(40.0000,-15.5000){\makebox(0,0)[rb]{$x_2-s=\frac{q}{p}(x_1-r)$}}%
\put(54.0000,-6.0000){\makebox(0,0)[rb]{$x_2=\frac{q}{p}x_1$}}%
\end{picture}%
 \caption{The unit square denotes the unit area of the two-torus. 
On its covering space, 
the line $x_2=\frac{q}{p}x_1$ expresses a cycle $S^1$ 
which winds $p$ times in $x_1$ direction and 
$q$ times in $x_2$ direction on the two-torus. 
This is the figure in the case of $(p, q)=(3, 2)$ and $(r, s)=(2, 1)$. 
Since $qr-ps=1$, the parallelogram spanned by $(p,q)$ and $(r, s)$ 
describes a unit area of another new two-torus. 
}
 \label{fig:f2}
\end{figure}

One can take other cycle $S^1$ which is transversal to the leaf of the 
foliation. 
They are characterized by relatively prime integers $p$ and $q$. 
The cycle is expressed by the line $x_2=q x_1/p$ 
on the covering space $\R^2$ of the two-torus. 
The periodicity of the 1-cycle is $p$ for $x_1$ direction 
and $q$ for $x_2$ direction (see Fig.\ref{fig:f2}). 
Let us introduce $r, s\in\Z$ which satisfy 
\begin{math}
 |\bps r & p\\ s & q\eps |=1
\end{math}\ .
The vectors $(x_1,x_2)=(r, s)$ and $(x_1,x_2)=(p, q)$ 
on the covering space then 
define the unit area of the torus. 
Fig.\ref{fig:f2} shows such a situation 
in the case $(p, q)=(3, 2)$ and $(r, s)=(2, 1)$. 
We have an ambiguity of the choice of $(r, s)$ such as 
\begin{equation}
 (r, s)\sim (r, s)+\Z(p, q)\ .
 \label{amb}
\end{equation}
However one can see later that this ambiguity has no matter for 
our problems. 
Let us consider the action $U_1$ on the cycle $S^1$ as in the previous 
subsection. 
To find the action, it is sufficient to get the point at which 
the leaf of the foliation $x_2=\theta x_1$ and 
the 1-cycle $(x_1, x_2)=(r+\lambda p, s+\lambda q)$, $\lambda\in\R$ 
intersect to each other. 
The intersecting point is then given by 
\begin{equation*}
 \lambda=\frac{r\theta-s}{q-p\theta}\ .
\end{equation*}
Here let $\theta':=\lambda$ and consider the functions on 
the cycle $S^1$ generated by $Z_2=e^{2\pi i z}$. On it 
$Z_2=e^{2\pi i z}$ acts by multiplication and 
$Z_1 : f(z)\mapsto f(z+\theta')$ acts as translation. 
$Z_1$ and $Z_2$ then have the relation 
\begin{equation}
 Z_1Z_2=e^{2\pi i\theta'}Z_2Z_1\ ,\qquad 
 \theta':= \frac{r\theta-s}{q-p\theta}\ .
 \label{morita}
\end{equation}
We denote by $\cA_{\theta'}$ the algebra generated by $Z_1$ and $Z_2$. 
Note that the ambiguity in Eq.(\ref{amb}) affects $\theta'$ as 
$\theta'-\Z$ so the relation (\ref{morita}) is independent of the 
ambiguity.

It is well-known that the noncommutativity of the Morita equivalent 
noncommutative tori are related by the fractional transformation 
as in Eq.(\ref{morita}). 
Generally, for a given foliation in a manifold $M$, 
one can define a $C^*$-algebra $\cA$ associated to it. 
Furthermore when one considers a closed submanifold $V$ of $M$ which 
is transversal to the leaf space of the foliation, 
it is known that $\cA$ is Morita equivalent to 
the algebra $\cA|_{V}$ which is obtained by `restricting' $\cA$ to $V$. 
Here $M$ is the two-torus, the foliation is the Kronecker foliation, 
$V$ is the cycle $S^1$ characterized by $(p, q)$, 
and the algebra $\cA|_{V}$ is the noncommutative two-torus $\cA_{\theta'}$ 
in the crossed product representation. 
Since the $C^*$-algebra $\cA$ and noncommutative torus $\cA_{\theta'}$ 
is Morita equivalent for any (relatively prime) integers $(p, q)$, 
we can say that $\cA_{\theta'}$ is Morita equivalent to $\cA_\theta$
\cite{co-book}. 

If we translate the torus by 
\begin{equation*}
 \bp x_1 \\ x_2\ep\raw \bp x_1' \\ x_2'\ep
  =\bp r & p \\ s & q\ep^{-1}\bp x_1 \\ x_2\ep
\end{equation*}
so that the vectors $(r, s)$ and $(p, q)$ are transformed to 
$(1, 0)$ and $(0, 1)$, respectively, 
the foliation on the torus is expressed as in Fig.\ref{fig:f2} 
but the slope is $\theta'$. 

In the next section, we relate the cycle $S^1$ to a D1-brane, 
whereas the leaf is identified with the orbit of the open string 
ending on the D1-brane in a certain limit.

\def \gh{{\hat g}}
\def \Bh{{\hat B}}
 \section{Open string ending on D1-brane}
\label{sec:3}

Let us consider a D1-brane winding on a two-torus. 
We denote its slope by $q/p$ and for simplicity 
we assume that it passes the origin. 
Now we introduce the metric on the two-torus as 
\begin{math}\gh=\bp \gh_{11} & \gh_{12}\\ \gh_{12} & \gh_{22}\ep\end{math}. 
The orbit of the open string with both ends on the D1-brane 
is determined so that the open string 
minimizes its length with respect to the metric $\gh$. 
Of course this determination is consistent 
with the first quantization picture of 
open strings\cite{K}. 

 \subsection{$(p, q)=(0, 1)$ case}
\label{ssec:31}

First, let us find the slope of the open string which minimizes its length 
in the simplest case $(p, q)=(0, 1)$. 
The situation is just the same as the argument in \cite{DH}, 
where the noncommutativity of the product of open string field is realized 
in a very intuitive and geometric way. 
The length square of the open string 
which winds once around $x_1$ direction is 
\begin{equation}
 \bp1 & \lambda\ep
 \bp \gh_{11} & \gh_{12}\\ \gh_{12} & \gh_{22}\ep
 \bp 1\\ \lambda\ep 
\end{equation}
where $\lambda$ is the slope of the orbit of the open string 
(see Fig.\ref{fig:f1}). 
Differentiating this length by $t$, one can easily see that 
it is minimized by $\lambda=-\gh_{12}/\gh_{22}$. 
Here set $\theta:=\lambda$ and one can relate this situation to that in 
Fig.\ref{fig:f1}. The D1-brane is the cycle $S^1$. 
The open string then winds along the leaf of the foliation. 

Let us consider field theory on the D1-brane. 
Note that the noncommutative product is not described by the 
Moyal product ($*$-product). 
We shall below explain that the noncommutative product structure 
can be identified with 
the crossed product defined in the previous section. 
The field corresponding to the open string ending on the D1-brane 
has two kinds of modes, the momentum along the D1-brane and 
the winding modes. Both modes are discretized and take their values in 
$\Z$. Let $m$ and $w$ be the momentum and the winding, respectively. 
The field $\phi$ can be Fourier-expanded by $m$ as 
\begin{equation*}
 \phi(x_2)=\sum_{m\in\Z}\phi_m U_2^m  
\end{equation*}
where $U_2=e^{2\pi i x_2}$. 
Furthermore $\phi_m$ is expanded by $w$ as 
$\phi_m=\sum_{w\in\Z}\phi_{w,m}U_1^w$ where $U_1$ is the generator 
defined in the previous section. 

Actually this expression of the field agrees with the open string physics 
as explained below. 
The product of the fields is identified with the open string interaction. 
When two open strings with modes $(w,m)$ and $(w',m')$ interact, 
an open string is created and its modes should simply be the sum of the two 
open string modes. 
Correspondingly, now the product of $U_1^wU_2^m$ with $U_1^{w'}U_2^{m'}$ 
in fact 
coincides with $U_1^{w+w'}U_2^{m+m'}$ up to a coefficient (phase factor). 
Moreover, the interaction of two open strings is defined 
by joining the end points of them. 
The open string with winding $w$ stretches with length $\theta w$ in 
$x_2$ direction. Therefore if the open string with modes $(w,m)=(1, 0)$ acts 
on the function on the D1-brane $f(x_2)$, 
the next open string acts on $f(x_2+\theta)$. 
This justifies the identification of the generator of the winding modes 
with $U_1$ defined in the previous section. 
The open string vertex is then described as the closed loop 
of the open strings. 
It in fact closes due to the 
momentum and winding modes conservation. 
The `integral' of the fields are then defined by 
\begin{equation*}
 \Tr \l(\sum_{w,m\in\Z^2}\phi_{w,m}U_1^w U_2^m\r)=\phi_{0,0}.
\end{equation*}
This is the standard definition of 
the `integral' on noncommutative two-tori. 

Thus, the field theory can be expressed as noncommutative two-tori.

 \subsection{General $(p, q)$ case}
\label{ssec:32}

Next we shall consider the D1-brane winding on general 1-cycles 
on the two-torus 
and discuss the relation to the Morita equivalence. 
We express the D1-brane as $x_2=q x_1/p$ where $p$ and $q$ are relatively 
prime integers (see Fig.\ref{fig:f2}). 
As in the previous subsection, we shall find the orbit of the 
open string ending on the D1-brane. 
We take $r, s\in\Z$ which satisfy 
\begin{math}
 |\bps r & p\\ s & q\eps |=1
\end{math}, 
and consider the open string ending on $(x_1, x_2)=(0, 0)$ and 
$(x_1, x_2)=(r+\lambda p, s+\lambda q)$. The latter is expressed as 
\begin{equation*}
 \bp x_1 \\ x_2\ep=\bp r+\lambda p\\ s+\lambda q \ep
 =\bp r & p \\ s & q\ep \bp 1\\ \lambda\ep.
\end{equation*}
Its length square with respect to the metric $\gh$ is then given by 
\begin{equation*}
 \bp r+\lambda p & s+\lambda q \ep\gh\bp r+\lambda p\\ s+\lambda q \ep
 =\bp1 & \lambda\ep
 \gh' 
 \bp 1\\ \lambda\ep 
\end{equation*}
where 
\begin{equation*}
 \gh'=\bp \gh'_{11} & \gh'_{12} \\ \gh'_{12} & \gh'_{22}\ep
 =\bp r & p \\ s & q\ep^t
  \bp \gh_{11} & \gh_{12} \\ \gh_{12} & \gh_{22}\ep
  \bp r & p \\ s & q\ep. 
\end{equation*}
Thus the argument reduces to the $(p, q)=(0, 1)$ case. 
The length of the open string is minimized at 
\begin{equation*}
 \lambda=-\frac{\gh'_{12}}{\gh'_{22}} 
\end{equation*}
and the slope of the orbit of the open string is then 
\begin{equation}
 \frac{s+\lambda q}{r+\lambda p}
 =-\frac{p\gh_{11}+q\gh_{12}}{p\gh_{12}+q\gh_{22}}.
 \label{moritaslope}
\end{equation}

We would like to relate this slope (\ref{moritaslope}) to 
that of the leaf of the foliation in Subsection \ref{ssec:22}. 
However, the slope (\ref{moritaslope}) depends on $p$ and $q$ 
in general $\gh$. 
Now let us consider the degenerate metric on the two-torus of the form 
\begin{equation}
 \gh=C\bp \theta^2 & -\theta\\ -\theta & 1\ep 
 \label{moritamet}
\end{equation}
where $C$ is a constant. 
Since the metric is degenerate, the length of the open string along the 
vector $(a, b)\in\R^2$ is the square root of 
\begin{equation*}
 \bp a & b\ep \gh \bp a \\ b\ep = C(a\theta-b)^2\ .
\end{equation*}
Clearly the mass of any open string stretching with slope $\theta$ is zero. 
In this case, the slope (\ref{moritaslope}) is independent 
of $(p, q)$ and coincides with $\theta$. 
Therefore the situation just coincides with that 
in Subsection \ref{ssec:22}. 
The open string field is 
represented as $\phi=\sum_{(w, m)\in\Z^2}\phi_{w, m}Z_1^wZ_2^m$. Here 
$Z_1$ and $Z_2$ are two generators of the noncommutative two-torus. 
They have the relation $Z_1Z_2=e^{2\pi i\theta'}Z_2Z_1$ where $\theta'$ is 
the one defined in Eq.(\ref{morita}). 

Thus we can conclude as follows. 
The two-torus with metric of the form in 
Eq.(\ref{moritamet}) is implicitly `foliated'. When we probe the 
physics on the two-torus by a D1-brane on a 1-cycle, the two-torus is 
`foliated' by the open strings ending on the D1-brane. 
Field theory of open strings on the two-torus is described by 
the noncommutative two-torus in the crossed product representation. 
Different choices of the 1-cycle on which the probe D1-brane lies 
are related by the Morita equivalence of the noncommutative two-tori. 

The above fact implies that the modular transformation $SL(2,\Z)$ 
preserves the form of the metric (\ref{moritamet}). 
Actually, 
\begin{equation}
 \bp r & p \\ s & q\ep^t
 \bp \theta^2 & -\theta \\ -\theta & 1 \ep
 \bp r & p \\ s & q\ep
 =(q-p\theta)^2
 \bp {\theta'}^2 & -\theta' \\ -\theta' & 1 \ep, 
 \label{thetap}
\end{equation}
where $\theta'=(r\theta-s)(q-p\theta)^{-1}$.

 \section{The Seiberg-Witten limit and the Kronecker foliation limit}
\label{sec:4}

\def \Eh{{\hat E}}

In this section, we discuss the meaning of the metric (\ref{moritamet}). 
Such forms of the metric are related to the so-called Seiberg-Witten 
limit\cite{SW}. 
The theory of D1-brane on the two-torus is obtained by T-dualizing 
the theory of D2-brane on the dual torus. 
Let $E:=g+B$ be the background (the pair of the metric $g$ and the $B$-field) 
on the torus where a D2-brane exists. 
The background $\Eh$ of the torus which is obtained by T-dualizing 
the D2-brane theory is then 
\begin{equation*}
 \Eh=(I_2 E+ I_1)(I_1 E + I_2)^{-1}\ ,\qquad 
 I_1:=\bp 1 & 0 \\ 0 & 0\ep,\ \ \ I_2:=\bp 0 &0  \\ 0 & 1\ep.
\end{equation*}
More explicitly, the background $\Eh:=\gh+{\hat B}$ is given by 
\begin{equation}
 \gh+{\hat B}
 =g_{11}^{-1}\bp 1 & B\\ B & \det{(g)}+B^2\ep 
 +g_{11}^{-1}\bp 0 & -g_{12}\\ g_{12} & 0\ep . 
 \label{Tdualbg}
\end{equation}
In the Seiberg-Witten limit $g\raw 0$, the above metric reduce to 
\begin{equation*}
 \gh\sim g_{11}^{-1}B^2
 \bp \ov{B^2} & \ov{B}\\
     \ov{B} & 1 \ep. 
\end{equation*}
This is nothing but the metric in Eq.(\ref{moritamet}) with 
$C=g_{11}^{-1}B^2$ and $\theta=-\ov{B}$. 
It is known that in the Seiberg-Witten limit 
the correlation function of open strings with both ends on D2-brane 
on a two-torus reduces to the algebra of the noncommutative torus\cite{SW}. 
The noncommutativity of the geometry in the Seiberg-Witten limit 
can be realized from the viewpoints of the deformation quantization. 
Alternatively, by T-dualizing one direction from a two-torus in 
the Seiberg-Witten limit, one obtains a two-torus which is degenerate 
for one direction. Such a geometry is directly related to the Kronecker 
foliation. 
D-branes are then transformed to D1-branes winding 
various direction. The physics of the open strings on the D1-brane 
can also be represented by the noncommutative torus, but in this side 
the noncommutativity is expressed in the crossed product. 
Thus, we can conclude that the two representations of 
the noncommutative two-tori, that in the deformation quantization and that 
in the crossed product, are related by T-dual.

 \section{Application}
\label{sec:ap}

On noncommutative tori, 
there exist explicit projective modules 
called Heisenberg modules\cite{KS}. 
Projective modules are noncommutative analogue of vector bundles. 
Therefore projective modules on two-tori correspond to 
noncommutative D2-branes (more precisely D2-D0 bound states). 
However the Heisenberg modules are constructed in the crossed product 
representation. Therefore they also have a D1-brane picture. 
In Subsection \ref{ssec:pm} we shall show that 
the Heisenberg modules can be defined 
from the viewpoint of D1-brane physics. 
Then in Subsection \ref{ssec:hms} we shall discuss the relation of 
these arguments to the homological mirror symmetry.

 \subsection{Projective modules from the Kronecker foliation}
\label{ssec:pm}

$\bullet$ {\it Heisenberg modules from D1-brane picture}

In Subsection \ref{ssec:21}, 
for a noncommutative two-torus $\cA_\theta$, 
a Morita equivalent noncommutative two-torus $\cA_{\theta'}$ was 
constructed geometrically. 
Two algebras $\cA$ and $\cA'$ are said to be Morita equivalent 
when there exists a projective module $E$ such that $\cA'\simeq \Ed{\cA}E$. 
Here $\Ed{\cA}E$ denotes the endomorphism algebra of $E$ which 
commutes with $\cA$. 
Such a projective module $E$ is then called a Morita equivalent 
bimodule. 
For noncommutative two-tori, 
Heisenberg modules are the Morita equivalent bimodule. 
It is used to prove that 
$\cA_\theta$ and $\cA_{\theta'}$ defined in Eq.(\ref{morita}) are 
Morita equivalent(see\cite{R,RS,KS}). 
Such a Heisenberg module over $\cA_\theta$ is 
characterized by $\cA_{\theta'}$ or equivalently 
\begin{math}
 \(\bps r & p\\s & q \eps\) \in SL(2,\Z)\ ,
\end{math}
so we denote it by $E_{q,p,\theta}$
\footnote{One can see later that actually the Heisenberg modules 
do not depend on the ambiguity of $(r, s)$ in Eq.(\ref{amb}). }. 
It is known that any finitely generated projective module is 
isomorphic to the direct sum of these Heisenberg modules
(see\cite{CR,KS}).

We shall construct the Heisenberg modules from the D1-brane picture below. 
As seen in Section \ref{sec:2}, the algebra $\cA_\theta$ is defined 
on the D1-brane of $(p,q)=(0,1)$ on the two-torus foliated with 
slope $\theta$. 
Thus we first fix the $S^1$ characterized 
by $(p,q)=(0,1)$ as the `base space'. 
\begin{figure}[h]
\unitlength 0.1in
\begin{picture}( 57.2500, 20.0000)(  4.7500,-32.0000)
%
\special{pn 8}%
\special{pa 1400 3000}%
\special{pa 6200 3000}%
\special{fp}%
%
\special{pn 8}%
\special{pa 1400 2200}%
\special{pa 6200 2200}%
\special{fp}%
%
\special{pn 8}%
\special{pa 1400 1400}%
\special{pa 6200 1400}%
\special{fp}%
%
\special{pn 8}%
\special{pa 3400 3200}%
\special{pa 3400 1200}%
\special{fp}%
%
\special{pn 8}%
\special{pa 4200 3200}%
\special{pa 4200 1200}%
\special{fp}%
%
\special{pn 8}%
\special{pa 2600 3200}%
\special{pa 2600 1200}%
\special{fp}%
%
\special{pn 8}%
\special{pa 5000 3200}%
\special{pa 5000 1200}%
\special{fp}%
%
\special{pn 8}%
\special{pa 2300 3200}%
\special{pa 5300 1200}%
\special{fp}%
%
\special{pn 8}%
\special{pa 2700 3200}%
\special{pa 5700 1200}%
\special{fp}%
%
\special{pn 8}%
\special{pa 3100 3200}%
\special{pa 6100 1200}%
\special{fp}%
%
\special{pn 8}%
\special{pa 1900 3200}%
\special{pa 4900 1200}%
\special{fp}%
%
\special{pn 8}%
\special{pa 1500 3200}%
\special{pa 4500 1200}%
\special{fp}%
%
\special{pn 8}%
\special{pa 3500 3200}%
\special{pa 6080 1480}%
\special{fp}%
%
\special{pn 8}%
\special{pa 4100 1200}%
\special{pa 1520 2920}%
\special{fp}%
%
\special{pn 8}%
\special{pa 3900 3200}%
\special{pa 6090 1740}%
\special{fp}%
%
\special{pn 8}%
\special{pa 3700 1200}%
\special{pa 1510 2660}%
\special{fp}%
%
\special{pn 8}%
\special{ar 1300 2200 100 100  3.1415927 4.7123890}%
%
\special{pn 8}%
\special{ar 1300 3000 100 100  1.5707963 3.1415927}%
%
\special{pn 8}%
\special{ar 1100 2500 100 100  6.2831853 6.2831853}%
\special{ar 1100 2500 100 100  0.0000000 1.5707963}%
%
\special{pn 8}%
\special{ar 1100 2700 100 100  4.7123890 6.2831853}%
%
\special{pn 8}%
\special{pa 1200 2200}%
\special{pa 1200 2500}%
\special{fp}%
%
\special{pn 8}%
\special{pa 1200 2700}%
\special{pa 1200 3000}%
\special{fp}%
\put(7.0000,-24.0000){\makebox(0,0){base}}%
\put(7.0000,-28.0000){\makebox(0,0){$S^1$}}%
%
\special{pn 20}%
\special{pa 1520 2920}%
\special{pa 2600 2200}%
\special{fp}%
%
\special{pn 20}%
\special{pa 2600 3000}%
\special{pa 3800 2200}%
\special{fp}%
%
\special{pn 20}%
\special{pa 3800 3000}%
\special{pa 5000 2200}%
\special{fp}%
%
\special{pn 20}%
\special{pa 1800 3000}%
\special{pa 3000 2200}%
\special{da 0.070}%
%
\special{pn 20}%
\special{pa 3000 3000}%
\special{pa 4200 2200}%
\special{da 0.070}%
%
\special{pn 20}%
\special{pa 4200 3000}%
\special{pa 5400 2200}%
\special{da 0.070}%
%
\special{pn 20}%
\special{pa 1510 2660}%
\special{pa 2200 2200}%
\special{dt 0.054}%
%
\special{pn 20}%
\special{pa 2200 3000}%
\special{pa 3400 2200}%
\special{dt 0.054}%
%
\special{pn 20}%
\special{pa 3400 3000}%
\special{pa 4600 2200}%
\special{dt 0.054}%
\end{picture}%
 \caption{A graphical viewpoint of the Heisenberg module in the case 
$(p,q)=(3, 2)$. There exist $p$ spirals over the base $S^1$. }
 \label{fig:vectbdl}
\end{figure}
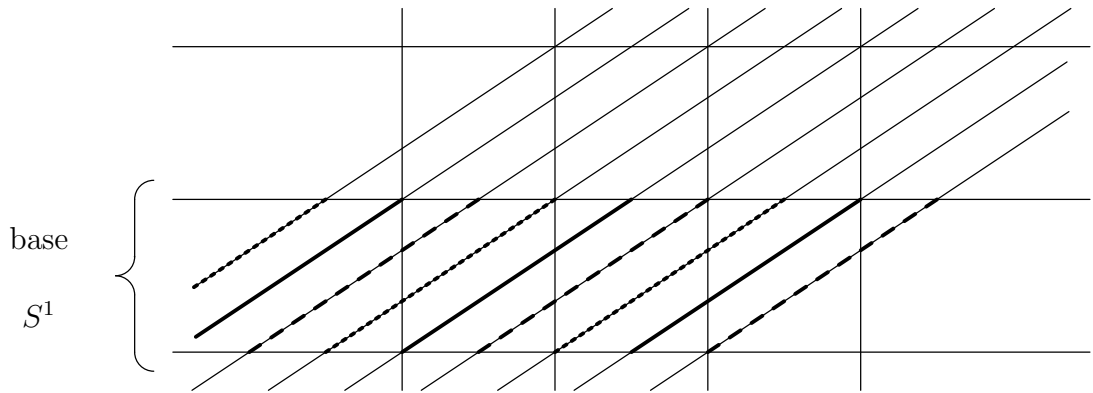
Let us construct module $E_{q,p,\theta}$ with relatively prime $p, q$. 
In this case, the endomorphism of the module 
$E_{q,p,\theta}$ is $\cA_{\theta'}$, 
that is, $\Ed{\cA_\theta}{E_{q,p,\theta}}$ is essentially that obtained 
in Subsection \ref{ssec:32} (Eq.(\ref{thetap})). 
Therefore we first consider the cycle characterized 
by $(p, q)$ on the covering 
space of the two-torus and regard it as lines over the base $S^1$. 
We have lines $x_1=px_2/q+l/q$ where $l\in\Z$ but, on the $S^1$, 
the line of $x_1$-axis $l/q$ coincides with that of $x_1$-axis $(l+p)/q$.  
Thus we have $p$ number of spirals $\R$ 
over the $S^1$ as in Fig.\ref{fig:vectbdl} 
and then consider a module of $p$ elements. 
We express it as 
\begin{equation*}
 f(z,\mu)\ ,\qquad z\in\R\, \ \ \  \mu=0,1,\cdots, p-1\ .
\end{equation*}
Here we identify the value $f(0,\mu)$ with the value of a function 
at $(x_1,x_2)=(0,\frac{\mu q}{p})$ on $(p,q)$ D1-brane. 
When we define the operation of $U_i$'s on this module, 
$z$ is regarded as the coordinate on the base $S^1$ with period $z\sim z+1$. 
Namely, the value of the function at point 
$(x_1,x_2)=(\frac{pz}{q}, z+\frac{\mu q}{p})$ 
is regarded as $f(z,\mu)$. 
$U_2$ is essentially the generator of functions on the $S^1$ 
as $U_2=e^{2\pi i z}$ in Subsection \ref{ssec:21}. 
However, now the action of $U_2$ should 
preserve the relation $f(z=\frac{q}{p},0)=f(0,\mu=1)$ because 
$f(\frac{q}{p},0)$ and $f(0,1)$ are the values of the function at the 
same points on the D1-brane. 
In this way the action of $U_2$ is determined as 
\begin{equation*}
  U_2 f(z,\mu)=f(z,\mu)e^{2\pi i(z+\mu\frac{q}{p})}\ .
\end{equation*}
The action of $U_1$ is also given in the same way as in Subsection 
\ref{ssec:21}, but we define it so that the open string shifts 
$\mu$ to $\mu+1$. The open string starting at $(x_1,x_2)=(0, z)$ then winds 
around $x_2$ direction and ends at $(x_1,x_2)=(0,z+\theta)$ 
on the base $S^1$, and $z+\theta$ is $z+\theta-\frac{q}{p}$ in the shifted 
coordinate. Thus $U_1$ is 
\begin{equation*}
  U_1 f(z,\mu)=f(z-\frac{q}{p}+\theta,\mu+1)\ .
\end{equation*}
On the other hand, $\cA_{\theta'}$ acts on the lines with slope $q/p$. 
The action of $Z_i$ is essentially that discussed 
in Subsection \ref{ssec:22}. 
However, due to the definition of the endomorphism, 
the action of $\cA_{\theta'}$ must commute with $U_1$ and $U_2$. 
It can be accomplished by identifying the coordinate on the lines 
of slope $q/p$ with the coordinate $z$ on the base $S^1$ ($x_1=0$) 
as follows. 
For each $\mu$, 
we consider the open string stretching between the line $x_1=0$ and 
the line $x_2=\frac{qx_1}{p}+(\frac{q\mu}{p}+\Z)$ on the covering space. 
The open string starting at $(x_1,x_2)=(0,z+\frac{q\mu}{p}+\Z)$ 
ends at point 
$(x_1,x_2)=(\frac{pz}{q-p\theta},\frac{qz}{q-p\theta}+\frac{q\mu}{p}+\Z)$. 
The corresponding coordinate on the line 
$x_2=\frac{qx_1}{p}+(\frac{q\mu}{p}+\Z)$ 
is then defined by $z$. 
\begin{figure}[h]
 \input{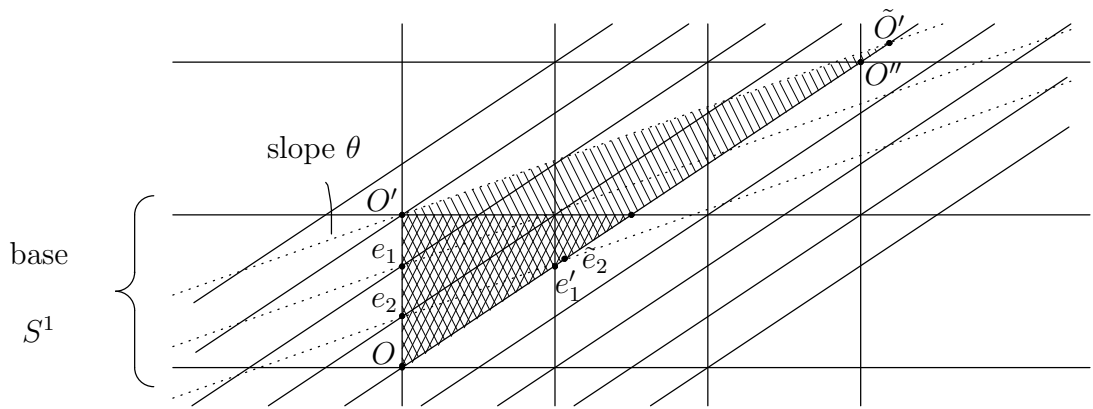}
 \caption{$U_i$ acts on cycle $x_1=0$ and $Z_i$ acts on 
the cycle characterized by $(p, q)$. 
For the action of $U_i$, $z$ in the expression of the module $f(z,\mu)$ 
can be regarded as the coordinate on $x_1=0$ with its periodicity 
$z\sim z+1$. In this figure, $|OO'|=1$. 
While, $z$ is regarded as the coordinate 
on $x_2=\frac{qx_1}{p}+(\frac{q\mu}{p}+\Z)$ 
when $Z_i$ acts on $f(z,\mu)$. 
Its periodicity is defined so that the action of $\cA_\theta$ commutes 
with that of $\cA_{\theta'}$. 
Then $|O{\tilde O}'|=1$ and the period is $|OO''|=q-p\theta$. }
 \label{fig:period}
\end{figure}

With this coordinate, the period of the D1-brane is $q-p\theta$ 
(see Fig.\ref{fig:period}). 
One can also see that $(z=(q-p\theta)\ov{p},\mu)$ and 
$(z=0,\mu+1)$ represent the same point on the D1-brane. 
For instance for $\mu=0$, $e_1'$ and $e_1$ in Fig.\ref{fig:period} 
is the same point. Thus, we define 
\begin{equation*}
  Z_2 f(z,\mu)=f(z,\mu)e^{2\pi i(\frac{z}{q-p\theta}+\frac{\mu}{p})}\ .
\end{equation*}
Moreover, an open string ending at a point $z$ on the D1-brane $\mu$ 
starts from the point $z-\ov{p}$ on the D1-brane $\mu+r$ 
up to $p\Z$. 
In Fig.\ref{fig:period} for $\mu=0$ the open string ending at ${\tilde e_2}$ 
starts from $e_2$. 
Therefore we define 
\begin{equation*}
 Z_1 f(z,\mu)=f(z-\ov{p},\mu+r)\ .
\end{equation*}
The action of $U_i$ and $Z_i$ for the same $i=1,2$ commutes trivially, 
and the action of $U_i$ and $Z_i$ for different $i$ commutes by 
the definition of the coordinate. 

To summarize, we have the following representation of 
the generators of algebra $\cA_\theta$ and $\cA_{\theta'}$
\footnote{This construction of Heisenberg modules is standard
(see \cite{KS}). 
However, the sign convention for $\mu$ is reversed 
compared to \cite{KS} 
so that the convention for Heisenberg modules agree with the one 
for an $A_\infty$-category in \cite{PZ}. See the next subsection. 
}. 
\begin{align}
 U_1 f(z,\mu)&=f(z-\frac{q}{p}+\theta,\mu+1)\\
 U_2 f(z,\mu)&=f(z,\mu)e^{2\pi i(z+\mu\frac{q}{p})}\\
 Z_1 f(z,\mu)&=f(z-\ov{p},\mu+r)\label{r}\\
 Z_2 f(z,\mu)&=f(z,\mu)e^{2\pi i(\frac{z}{q-p\theta}+\frac{\mu}{p})}
\end{align}
In this expression, $U_i$ and $Z_i$ satisfy the following relation 
\begin{equation*}
 U_1U_2=e^{2\pi i\theta}U_2U_1\ ,\qquad 
 Z_1Z_2=e^{2\pi i(-\theta')}Z_2Z_1\ .
\end{equation*}
We define on this Heisenberg module the action of $U_i$ as left 
action and that of $Z_i$ as right action. 
Therefore, the sign of $-\theta'$ flips. 
That is, this module is regarded as a $\cA_\theta$-$\cA_{\theta'}$ bimodule. 
As mentioned in Eq.(\ref{amb}), the choice of 
$(r, s)$ has an ambiguity when relatively prime integers $p, q$ are given. 
However one can confirm from Eq.(\ref{r}) 
that the module $E_{q,p,\theta}$ does not depend 
on the ambiguity since $\mu$ is defined up to $p\Z$.

It is known that the module obtained above equips the following 
constant curvature connection. 
\begin{equation}
 \nabla_1=\frac{2\pi i p}{q-p\theta}z\ ,\qquad 
 \nabla_2=\fpartial{z}\ .
 \label{ccc}
\end{equation}
Thus, modules $E_{q,p,\theta}$ has a constant curvature 
$[\nabla_1,\nabla_2]=\frac{2\pi i p}{q-p\theta}$.

\begin{figure}[h]
\unitlength 0.1in
\begin{picture}( 52.0000, 18.0000)( 12.0000,-34.0000)
%
\special{pn 8}%
\special{pa 1200 3200}%
\special{pa 6400 3200}%
\special{fp}%
%
\special{pn 8}%
\special{pa 1200 2000}%
\special{pa 6400 2000}%
\special{fp}%
%
\special{pn 8}%
\special{pa 3200 3400}%
\special{pa 3200 1600}%
\special{fp}%
%
\special{pn 8}%
\special{pa 4400 3400}%
\special{pa 4400 1600}%
\special{fp}%
%
\special{pn 8}%
\special{pa 5600 3400}%
\special{pa 5600 1600}%
\special{fp}%
%
\special{pn 8}%
\special{pa 2000 3400}%
\special{pa 2000 1600}%
\special{fp}%
%
\special{pn 13}%
\special{pa 2000 3200}%
\special{pa 2000 2000}%
\special{fp}%
%
\special{pn 13}%
\special{pa 2000 3200}%
\special{pa 3200 2000}%
\special{fp}%
%
\special{pn 13}%
\special{pa 2000 3200}%
\special{pa 5600 2000}%
\special{fp}%
\put(23.0000,-24.0000){\makebox(0,0){$E_{q-p\theta}$}}%
\put(29.0000,-26.0000){\makebox(0,0){$E_{q'-p'\theta'}$}}%
%
\special{pn 4}%
\special{ar 2000 3200 600 600  4.7123890 5.4977871}%
%
\special{pn 4}%
\special{ar 2000 3200 800 800  5.4977871 5.9614348}%
%
\special{pn 8}%
\special{ar 2000 3200 400 400  4.7123890 5.9614348}%
%
\special{pn 8}%
\special{pa 2290 3010}%
\special{pa 2278 3040}%
\special{pa 2266 3070}%
\special{pa 2254 3100}%
\special{pa 2246 3132}%
\special{pa 2240 3162}%
\special{pa 2238 3194}%
\special{pa 2238 3226}%
\special{pa 2238 3258}%
\special{pa 2240 3290}%
\special{pa 2240 3290}%
\special{sp}%
\put(22.5000,-34.5000){\makebox(0,0){$E_{Q-P\theta}$}}%
\put(31.0000,-21.1000){\makebox(0,0)[lt]{slope $\frac{q}{p}$}}%
\put(44.3000,-24.0000){\makebox(0,0)[lt]{slope $\frac{Q}{P}$}}%
\end{picture}%
 \caption{The graphical picture of the tensor product in the case of 
$(p, q)=(1, 1)$ and $(p', q')=(1, 2)$ ($(P, Q)=(1, 3)$). 
Each module is characterized by the pair of rational slopes of lines. }
 \label{fig:tt}
\end{figure}
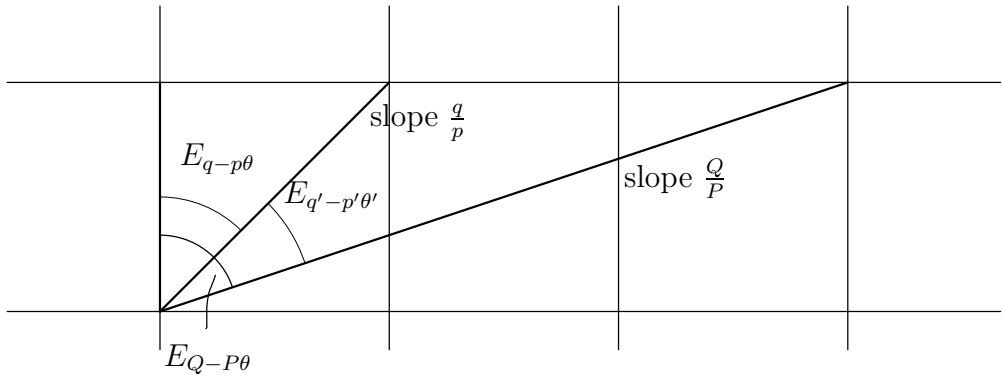

The Heisenberg modules constructed above are regarded as that defined on 
the base space $S^1$ characterized by $(0,1)$. 
These can be generalized easily in 
the case that $(0, 1)$ is replaced to any (relatively prime) 
$(p, q)$. In this case the modules are characterized by two cycles 
$x_2=\frac{q}{p}x_1$ and $x_2=\frac{Q}{P}x_1$. 
The corresponding module can be constructed in a quite similar way as 
the $(p, q)=(0, 1)$ case above. 
Consequently one obtains Heisenberg module 
$E_{q',p',\theta'}$ over $\cA_{\theta'}$ where 
\begin{equation*}
 \bp R& P\\ S& Q\ep=\bp r& p\\ s& q\ep \bp r'& p'\\ s'& q'\ep\ ,
\end{equation*}
or equivalently 
$\cA_{\theta'}$-$\cA_{\theta''}$ bimodules where 
\begin{math}
 \theta'':=\frac{R\theta-S}{Q-P\theta}
\end{math}
(see Fig.\ref{fig:tt}).

\vspace*{0.2cm}

$\bullet$ {\it The tensor product}

The module $E_{q,p,\theta}$ constructed above can also be regarded as 
a $\cA_{\theta}$-$\cA_{\theta'}$ bimodule. 
More generally, as mentioned above, a module characterized by the pair 
$x_1=px_2/q$ and $x_1=Px_2/Q$ belongs to 
a $\cA_{\theta'}$-$\cA_{\theta''}$ bimodule. 
Note that bimodules are regarded as open strings\cite{SW,KMT}. 
The line $x_1=px_2/q$ (resp. $x_1=Px_2/Q$) characterizes 
module $E_{p,q,\theta}$ (resp. $E_{P,Q,\theta}$) over $\cA_\theta$ 
by fixing the base $S^1$ as $x_1=0$. 
The module $E_{p,q,\theta}$ corresponds to the bound state of 
$p$ D2-branes and $-q$ D0-branes and 
the module $E_{P-Q\theta}$ also does similarly. 
The $\cA_{\theta'}$-$\cA_{\theta''}$ bimodule is then 
regarded as the open string stretching between the D-branes 
$E_{p,q,\theta}$ and $E_{P,Q,\theta}$. 

Now, let us consider the D-branes represented by the lines 
$x_1=0$, $x_1=px_2/q$ and $x_1=Px_2/Q$. We have 
the open strings which belong to 
a $\cA_{\theta}$-$\cA_{\theta'}$ bimodule $E_{q,p,\theta}$ and 
a $\cA_{\theta'}$-$\cA_{\theta''}$ bimodule $E_{q',p',\theta'}$. 
These open strings interact on the D1-brane $x_1=px_2/q$ 
and then become an open string stretching between 
the D1-brane $x_1=0$ and $x_1=Px_2/Q$. 
Then the open string should be expressed as an element of 
a $\cA_{\theta}$-$\cA_{\theta''}$ bimodule $E_{Q,P,\theta}$.

Such open string interactions can be accomplished by the 
tensor product between bimodules. 
The tensor product between $E_{q,p,\theta}$ and $E_{q',p',\theta'}$ 
is defined 
so that $E_{q,p,\theta}a\otimes_{\cA_{\theta'}} E_{q',p',\theta'}\sim
E_{q,p,\theta}\otimes_{\cA_{\theta'}}a E_{q',p',\theta'}$ for 
$a\in\cA_{\theta'}$. 
Fig.\ref{fig:tt} shows a geometric picture of the tensor product 
in the case 
\begin{math}
 \l(\bps r& p\\ s& q\eps\r)=\l(\bps 1 & 1\\ 0 & 1\eps\r) 
\end{math}
and 
\begin{math}
 \l(\bps r'& p'\\ s'& q'\eps\r)=\l(\bps 1 & 2\\ 0 & 1\eps\r) 
\end{math}. 
Let $\varphi$ be the isomorphism from 
$E_{q,p,\theta}\otimes_{\cA_{\theta'}} E_{q',p',\theta'}$ 
to $E_{Q,P,\theta}$. 
Such an isomorphism is recently constructed explicitly in \cite{Stensor}
\footnote{Precisely, the tensor product constructed in \cite{Stensor} 
is different from that presented here. 
In \cite{Stensor} the tensor product between right $\cA_\theta$ modules 
and left $\cA_\theta$ modules is constructed. 
In that case, for right $\cA_\theta$ module $E_{q,p,\theta}$ and 
left $\cA_\theta$ modules $E_{q',p',\theta}$, 
the tensor product is defined so that 
$E_{q,p,\theta}U_i\otimes_{\cA_\theta}E_{q',p',\theta}
\sim E_{q,p,\theta}\otimes_{\cA_\theta}U_iE_{q',p',\theta}$ 
where $U_i$'s are two generators of $\cA_\theta$. 
We instead define the tensor product 
as in Eq.(\ref{contract}) for our purposes. }.
It is given by 
\begin{equation}
 \varphi(f\otimes g)(z,\rho)=
 \sum_{u\in\Z}f(z+\ov{p}(u-\frac{p'}{P}\rho),-ru+\rho)\cdot
 g(\frac{z}{q-p\theta}-\frac{q'-p'\theta'}{p'}(u-\frac{p'}{P}\rho), u)\ .
 \label{tensorprod}
\end{equation}
Actually, one can check that this tensor product satisfies 
\begin{equation}
 \varphi((Z_i f)\otimes g)=\varphi(f\otimes (U_i g))
 \label{contract}
\end{equation}
for $i=1, 2$. 
Moreover it is defined so that 
\begin{equation*}
 \begin{split}
  \varphi((U_i f)\otimes g)&=U_i\varphi(f\otimes g)\\
  \varphi(f\otimes (Z_i g))&=Z_i\varphi(f\otimes g)
 \end{split}
\end{equation*}
for $i=1, 2$. 
In the above equations, 
$U_i$'s in the left hand side are those which 
act on $E_{q,p,\theta}$ and $U_i$'s in the right hand side are 
those which act on $E_{Q,P,\theta}$. Similarly $Z_i$'s in both 
sides are different from each other. 

\def \Mh{{\hat M}}
\def \P{{\cal P}}
\def \Q{{\cal Q}}

 \subsection{Relation to the homological mirror symmetry}
\label{ssec:hms}

It is known that the two-torus with background $E=g+B$ is mirror dual 
to the T-dual torus which has background $\Eh=\gh+\Bh$ (Eq.(\ref{Tdualbg})). 
Mirror symmetry is a symmetry between Calabi-Yau manifolds. 
It is related to a closed string physics as well as 
the T-duality is. 
Instead, one can also consider D-branes on the manifolds 
and the (topological) open strings between them\cite{W2}. 
These tools may enable us to realize the symmetry between 
mirror manifolds. 
Homological mirror symmetry proposal\cite{mirror} is a 
mathematical setup of such an attempt. 
On Calabi-Yau manifolds 
one can define two categories 
which are related to the topological open string (A-model or B-model) 
in \cite{W2}. 
Then the conjecture is that 
a category on a Calabi-Yau manifold $M$ is equivalent to another 
category on the mirror dual Calabi-Yau manifold $\Mh$. 
Note that a category consists of objects and morphisms between the objects, 
where the composition of morphisms satisfies the associativity. 
One can identify the objects as some kind of D-branes and 
the morphisms as open strings between them. 
The category related to the A-model is an $A_\infty$-category
\footnote{An $A_\infty$-category is a generalization of the usual category. 
It consists of objects and morphisms, but generally the composition 
of the morphisms is not associative. Instead, a $A_\infty$-category has 
$i$-linear maps $m_i$ of the morphisms for $i\ge 1$ and they satisfy 
certain relations. The relations contain a condition for $m_2$, and 
it reduces to the associativity relation of $m_2$ when $m_1=0$. 
The $A_\infty$-category which we shall consider is just in this case, 
so it is also the usual category with $m_2$ the composition 
of the morphisms. 
}. Essentially it is constructed by (pseudo) holomorphic map of 
open string disks. 
On the other hand, the category related to the B-model is 
the derived category
\footnote{For the definition of the derived category, see 
\cite{GM}. A brief introduction of it for physicists, 
see the review part of \cite{DJP} for example. 
This paper does not need the precise definition. } 
of coherent sheaves. 
This side is better understood physically \cite{Dou,AL,Dia}. 
It is then conjectured that the bounded derived category of 
coherent sheaves on $M$ is equivalent to the bounded derived category 
of a suitable $A_\infty$-category 
\footnote{In fact, it is still unknown precisely that 
what kind of $A_\infty$-category should correspond to the 
bounded derived category of coherent sheaves for general 
(Calabi-Yau) manifolds. 
An $A_\infty$-category is constructed in \cite{Fukaya} 
and refined in \cite{mirror} to propose the homological mirror 
symmetry. 
However generally, 
the bounded derived category of coherent sheaves is `larger'. 
It is then believed that some modification of the $A_\infty$-category 
in \cite{mirror} corresponds to the 
bounded derived category of coherent sheaves. 
However, in the case of two-tori which we shall discuss below, 
this problem is no matter. 
}
and vice visa. 
In two-tori case, this statement is proved in \cite{PZ}. 
We shall avoid explaining the precise statements, and review shortly a part 
of the story 
which is relevant here. 
For the two-torus with background $E=g+B$, the complex structure $\tau$ and 
the complexified symplectic structure $\rho$ are defined as\cite{DVV}
\begin{equation}
 \tau=\frac{g_{12}}{g_{11}}+i\frac{\sqrt{g}}{g_{11}}
\ ,\qquad \rho=i\sqrt{g}+B\ .
 \label{taurho}
\end{equation}
Thus, on two-tori, the pairs of the complex structure and 
the complexified symplectic structure are in one-to-one correspondence 
with the backgrounds $E=g+B$. 
The mirror dual torus is defined so that 
the complex structure and the complexified symplectic structure 
are interchanged, that is, ${\hat \tau}=\rho$ and ${\hat \rho}=\tau$ 
where ${\hat \tau}$ and ${\hat \rho}$ are the complex structure 
and the complexified symplectic structure, respectively, 
on the mirror dual torus. 
One can see that the complex and the complexified symplectic structure 
on the mirror dual torus (${\hat \tau}$ and ${\hat \rho}$, respectively) 
are nothing but those 
defined by the T-dual background $\Eh=\gh+{\hat B}$ in Eq.(\ref{Tdualbg}). 

Let us consider the two categories; 
the derived category of coherent sheaves is defined by using the 
complex structure and is independent of the complexified symplectic 
structure, while the $A_\infty$-category depends 
only on the complexified symplectic structure. 
Then the arguments of \cite{PZ} are as follows. 
For the bounded derived category of coherent sheaves 
on the two-torus with complex structure $\tau$, 
holomorphic vector bundles (and skyscraper sheaf) are 
considered as the objects and 
homomorphisms between them are then constructed as the morphisms. 
It is known that holomorphic sections of vector bundles on tori are
described by theta functions. 
These holomorphic sections are homomorphisms between 
trivial bundles and the corresponding vector bundles, 
and in fact any other homomorphism is also described by theta functions. 
Between these homomorphisms expressed by theta functions, 
there is a product structure which is given simply by multiplication 
of two functions on the covering space of the two-torus. 
On the other hand, on the dual torus with 
complexified symplectic structure ${\hat \rho}(=\tau)$, 
an $A_\infty$-category is defined as follows. 
An object of the $A_\infty$-category is a pair of 
a special lagrangian submanifold and a flat bundle on it. 
Here special lagrangian submanifolds on the two-torus are given 
by geodesic cycles $S^1$, \ie the geodesic lines on the covering space. 
If the flat bundle is a line bundle, the object is just the 
D1-brane discussed until now. An object with rank $n$ flat bundle 
is then $n$ D1-branes on a cycle $S^1$. 
The morphisms of the $A_\infty$-category are then the homomorphism between 
two flat bundles on each geodesic cycle $S^1$. 
Thus, they are regarded as open strings. 
Two open strings interact when their endpoints belong to the same 
D-branes just as the tensor product in the previous 
subsection. Correspondingly, a product structure $m_2$ is defined 
in the $A_\infty$-category. Essentially it is defined by summing up 
holomorphic disks
\footnote{Note that, however, the product structure does not depend on 
the holomorphic structure ${\hat \tau}=\rho$ and defined 
by using ${\hat \rho}=\tau$.}
(see later arguments). 
It is then proved in \cite{PZ} that the derived category of coherent 
sheaves and the $A_\infty$-category are equivalent as categories. 
This means, between these two categories, 
there exists a functor which induces both the isomorphism 
of objects and that of morphisms. 
In particular, the isomorphism between morphisms, 
\ie the isomorphism between theta functions and 
the morphisms in the $A_\infty$-category, 
is compatible with the product structures on both sides. 
This implies that the product $m_2$ on the dual torus is compatible 
with the addition formula of theta functions 
on the torus with $\tau$ as mentioned in \cite{mirror}. 

This is the correspondence of the two categories 
attending to $\tau={\hat \rho}$. 
Of course, one can discuss the correspondence 
attending to $\rho={\hat \tau}$ in the same way. 

Next, we shall argue the relation of the above arguments to noncommutative 
tori. The argument above holds for general values of $\tau$, \ie 
the two-tori with general flat backgrounds. 
Let us consider the Seiberg-Witten limit in this situation. 
It corresponds to the limit $\Im\rho\raw 0$. 
Because $\rho={\hat \tau}$, 
the complex structure of the dual torus is degenerate 
in this limit. Such tori are expressed in terms of noncommutative 
tori as argued previously. 
Namely, for such a degenerate two-torus one can define a noncommutative 
torus $U_1U_2=e^{-2\pi i\ov{\hat\tau}}U_2U_1$ where 
$-\ov{\hat \tau}=-\ov{B}=\theta$
\footnote{The noncommutativity is actually $e^{-2\pi i\ov{\hat\tau}}$ 
since we define the complex structure as in Eq.(\ref{taurho}) and 
the noncommutative torus as in Subsection \ref{ssec:31}. 
We can also define the noncommutativity of the two-torus as 
$e^{2\pi i{\hat\tau}}$ by changing these definitions relatively.
}. 
In this direction, the noncommutative compactification of 
module space of complex structures ${\hat \tau}$ for elliptic curves 
is discussed in \cite{So,SV}. 
On the other hand, for the moduli parameter $\tau={\hat \rho}$ 
the degeneration is irrelevant and the correspondence of the two categories 
explained above remains true in this limit. 
We shall, however, argue the situation related to the categories 
associated to $\tau={\hat \rho}$ side below. 

\vspace*{0.2cm}

$\bullet$ {\it A noncommutative homological mirror symmetry}

As stated previously, projective modules are regarded as noncommutative 
analogue of vector bundles. 
Actually, for a Heisenberg module $E_{q,p,\theta}$ with $q-p\theta>0$, 
$p$ denotes the first Chern class of the projective module. 
$q$ then coincides with the rank of the projective module 
in the commutative case. 
Generally in noncommutative case, the rank is defined as $q-p\theta$, 
which is just identified with the value proportional to 
the mass of the $q$ D2- $-p$ D0 brane 
bound state\cite{BKMT,KMT,K}.)
Moreover, as stated in the 
previous subsection, the Heisenberg modules can also be regarded as 
bimodules. Note that when one regards Heisenberg modules 
$E_{q,p,\theta}$ and 
$E_{Q,P,\theta}$ as noncommutative vector bundles over $\cA_\theta$, 
the $\cA_{\theta'}$-$\cA_{\theta''}$ bimodule, 
which is also constructed as a Heisenberg module, 
is regarded as $\mathrm{Hom}(E_{q,p,\theta},E_{Q,P,\theta})$. 
Namely, these bimodules are regarded as noncommutative analogue of morphisms 
of the (derived) category of the coherent sheaves 
if a noncommutative analogue of the holomorphic structure is defined. 
On the other hand, in the previous subsection we construct the Heisenberg 
modules in the D1-brane picture. 
Since the D1-branes are lagrangian submanifolds, the Heisenberg modules 
are also expected to have a geometric picture of the $A_\infty$-category. 
Now on the noncommutative torus we consider an additional structure 
$\tau\in \C$ which is introduced as a 
noncommutative analogue of the holomorphic structure in \cite{Stensor} 
(for higher dimensional tori, 
see \cite{Stheta}). 
We shall then consider holomorphic submodules of the Heisenberg modules 
and show that the tensor product is equivalent to the 
$A_\infty$-structure $m_2$ on the dual torus (with background $\Eh$). 
Namely, the tensor product is realized as summing up the holomorphic disks. 
Since the definition of the $A_\infty$-category is complicated 
even for two-tori, 
we avoid to write down the definition. 
For the readers who would like to know the precise definition, see 
\cite{PZ}. 
We shall first take the tensor product and then 
rewrite it to the form of $m_2$ of the $A_\infty$-category. 
Categorically, the result implies the compatibility of the 
product structures. The correspondence of objects and that of morphisms 
are more clear (compare the arguments below with\cite{PZ}).

We have seen that the Heisenberg modules equip the constant 
curvature connection as in Eq.(\ref{ccc}). It is known that 
all constant curvature connections on $E_{q,p,\theta}$ are given as 
\begin{equation*}
  \nabla_1=\frac{2\pi i p}{q-p\theta}z+2\pi i c_1\ ,\qquad 
 \nabla_2=\fpartial{z}+2\pi i c_2
\end{equation*}
where $c_1, c_2$ are real numbers\cite{CR,KS}. Then consider the solution 
of the following equation
\begin{equation*}
 (-\tau\nabla_1+\nabla_2)f(z,\mu)=0\ .
\end{equation*}
Its solutions are written in the form 
\begin{equation}
 f(z,\mu)=a\exp{\(\pi i\tau \frac{p}{q-p\theta}z^2
 +2\pi i(-\tau c_1+c_2)z\)}\ ,\qquad a\in \C^p\ .
 \label{holvect}
\end{equation}
$-\tau\nabla_1+\nabla_2$ is regarded as the noncommutative analogue of 
the Dolbeault operator $\bar\partial$, 
and the solutions are called holomorphic vectors\cite{Stensor}. 
Thus, for each Heisenberg module, 
one can consider its submodule over $\cA_\theta$. 
The elements are expressed as linear combinations of 
the holomorphic vectors. 
The holomorphic vectors are labeled by the continuous parameter 
$-\tau c_1+c_2\in\C$, so the submodule is not so small. 
Moreover, the tensor product of two holomorphic vectors is also 
a holomorphic vector\cite{Stensor}. 

Now we express $a=a^\mu e_\mu$ where $e_\mu$ is the basis of the modules of 
$\mu$ and rewrite Eq.(\ref{holvect}) in the following form
\begin{equation}
 f(z)=a^\mu e_\mu \exp{\(\pi i\tau \frac{p}{q-p\theta}
(z-\frac{q-p\theta}{p}\epsilon)^2\)} 
 \label{holvect2}
\end{equation}
where $\epsilon=c_1-c_2/\tau$. 
The tensor product is then written explicitly as 
\begin{align}
& \varphi(e_\mu\exp{(\pi i\tau \frac{p}{q-p\theta}
(z-\frac{q-p\theta}{p}\epsilon)^2)}\otimes
e_\nu\exp{(\pi i\tau \frac{p'}{q'-p'\theta'}
(z-\frac{q'-p'\theta'}{p'}\epsilon')^2)}) \nonumber\\
&\qquad=c_{\mu\nu}^\rho e_\rho
\exp{(\pi i\tau \frac{P}{Q-P\theta}(z-\frac{Q-P\theta}{P}
(\epsilon+\frac{1}{q-p\theta}\epsilon'))^2)}\ , \label{holtensor}\\
&c_{\mu\nu}^\rho=\sum_{u\in\Z}\delta_\mu^{-ru+\rho}\delta_\nu^u
\exp{\(\frac{\pi i\tau}{pp'P}
(Pu-p'\rho+p(q'-p'\theta')\epsilon'-p'\epsilon)^2\)}\ .\label{strconst}
\end{align}
Thus one can see that actually the holomorphic vectors close with respect 
to the tensor product. 
Moreover, this tensor product is associative. 
For $f\sim \exp{\(\pi i\tau \frac{p}{q-p\theta}
(z-\frac{q-p\theta}{p}\epsilon)^2\)}\in E_{q,p,\theta}$, 
$g\sim \exp{\(\pi i\tau \frac{p'}{q'-p'\theta'}
(z-\frac{q'-p'\theta'}{p'}\epsilon')^2\)}\in E_{q',p',\theta'}$ and 
$h\sim \exp{\(\pi i\tau \frac{p''}{q''-p''\theta''}
(z-\frac{q''-p''\theta''}{p''}\epsilon'')^2\)}\in E_{q'',p'',\theta''}$, 
$\varphi(f\otimes g\otimes h)$ becomes 
$\exp{\(\pi i\tau \frac{\P}{\Q-\P\theta}
(z-\frac{\Q-\P\theta}{\P}(\epsilon+\frac{\epsilon'}{q-p\theta}
+\frac{\epsilon''}{Q-P\theta}))^2\)}$ with an appropriate 
coefficient where 
\begin{equation*}
 \bp R & P\\ S & Q\ep=\bp r & p\\ s & q\ep
 \bp r' & p'\\ s' & q'\ep\ ,\qquad 
 \bp {\cal R} & \P \\ {\cal S} & \Q\ep=\bp R & P\\ S& Q\ep
 \bp r'' & p''\\ s'' & q''\ep\ . 
\end{equation*}

\begin{figure}[h]
 \input{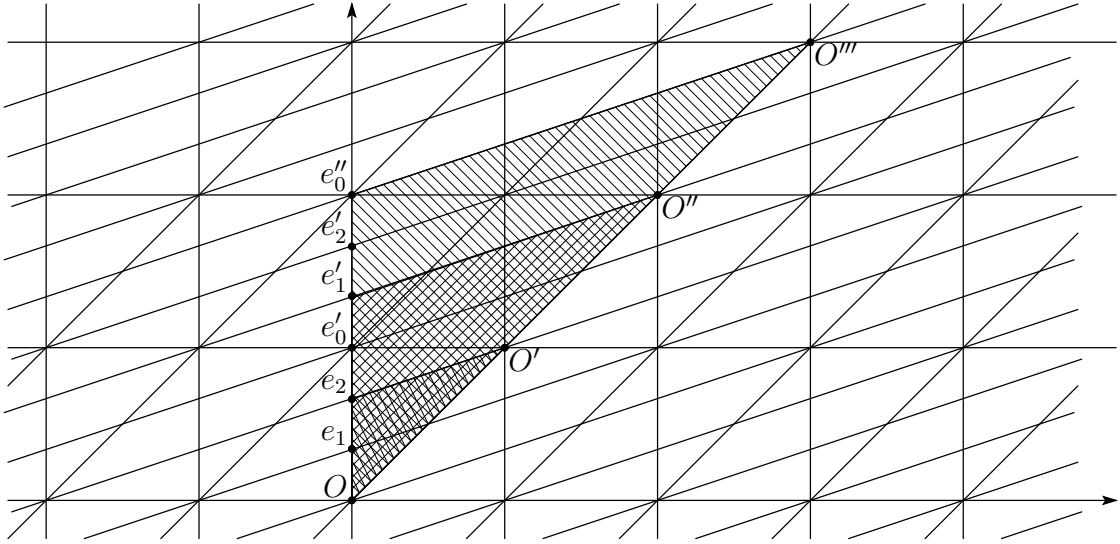}
 \caption{The area of the triangle times the complexified symplectic form 
gives the coefficient of the tensor product. 
This is the figure in the case 
$(p,q)=(1,1)$, $(P, Q)=(3, 1)$ ($(p',q')=(2, 1)$), $\epsilon_i=0$ 
and $\nu=0$ (since $p=1$, $\mu=0$ originally). 
$e_{\mu=0}$, $e_{\nu=0}$ and $e_{\rho=0}$ correspond to the point $O$. 
$'$ means the mirror points (in the sense of the covering) 
of the original point. 
$e_i$s in this picture express $e_{\rho=i}$. 
The triangle $OO'e_2$, $OO''e_1'$ and $OO'''e_0''$ correspond to 
$\triangle_{m=1}$, $\triangle_{m=2}$ and $\triangle_{m=3}$, 
respectively. 
}
 \label{fig:tri}
\end{figure}

We have a geometric realization about $\epsilon$ in holomorphic vectors 
and the structure constant $c_{\mu\nu}^\rho$ in Eq.(\ref{strconst}). 
Such a realization will enable us to understand the relation 
between this tensor product and the $A_\infty$-category on two-tori. 
First, let us rewrite $\epsilon$ and $\epsilon'$ as 
\begin{equation*}
 \epsilon:=\frac{q}{q-p\theta}\epsilon_2-\epsilon_1\ ,\qquad 
 \epsilon':=(q-p\theta)\(\frac{Q}{Q-P\theta}\epsilon_3-
 \frac{q}{q-p\theta}\epsilon_2\)\ .
\end{equation*}
The structure constant $c_{\mu\nu}^\rho$ is then expressed as 
\begin{equation}
 c_{\mu\nu}^\rho:=\sum_{m\in\Z}\delta_{R(p'm+Q\mu+q'\nu)}^\rho
 \exp{\(2\pi i\tau\triangle_m\)}
 \label{strconst2}
\end{equation}
where $\triangle_m$ is of the form 
\begin{equation}
 \begin{split}
 \triangle_m
&=\ov{2pp'P}
 \(pp'm+pq'\nu-p'\mu+pQ\epsilon_3+\epsilon_1p'
 -Pq\epsilon_2\)^2\\
&=\ov{2pp'P}
 \(pp'm+pq'\nu-p'\mu+pQ(\epsilon_3-\epsilon_2)-p'(\epsilon_2-\epsilon_1)
\)^2\ .
 \end{split}
 \label{tri}
\end{equation}
\begin{figure}[h]
\unitlength 0.1in
\begin{picture}( 54.4000, 20.0000)(  7.6000,-32.0000)
%
\special{pn 8}%
\special{pa 1200 2800}%
\special{pa 6200 2800}%
\special{fp}%
\special{sh 1}%
\special{pa 6200 2800}%
\special{pa 6134 2780}%
\special{pa 6148 2800}%
\special{pa 6134 2820}%
\special{pa 6200 2800}%
\special{fp}%
%
\special{pn 8}%
\special{pa 3800 3200}%
\special{pa 3800 1200}%
\special{fp}%
\special{sh 1}%
\special{pa 3800 1200}%
\special{pa 3780 1268}%
\special{pa 3800 1254}%
\special{pa 3820 1268}%
\special{pa 3800 1200}%
\special{fp}%
%
\special{pn 8}%
\special{pa 3100 3200}%
\special{pa 5100 1200}%
\special{fp}%
%
\special{pn 8}%
\special{pa 1200 3000}%
\special{pa 6180 1340}%
\special{fp}%
%
\special{pn 8}%
\special{pa 4080 2040}%
\special{pa 3800 2320}%
\special{fp}%
\special{pa 4170 2010}%
\special{pa 3800 2380}%
\special{fp}%
\special{pa 4260 1980}%
\special{pa 3800 2440}%
\special{fp}%
\special{pa 3990 2070}%
\special{pa 3800 2260}%
\special{fp}%
\special{pa 3900 2100}%
\special{pa 3800 2200}%
\special{fp}%
\put(34.5000,-27.5000){\makebox(0,0)[rb]{$-\alpha_2$}}%
\put(17.5000,-27.5000){\makebox(0,0)[rb]{$-\alpha_3$}}%
\put(61.0000,-29.5000){\makebox(0,0){$x_1=:y$}}%
\put(37.0000,-14.0000){\makebox(0,0)[rb]{$x_2=:x$}}%
%
\special{pn 13}%
\special{sh 1}%
\special{ar 3800 2130 10 10 0  6.28318530717959E+0000}%
\special{sh 1}%
\special{ar 3800 2130 10 10 0  6.28318530717959E+0000}%
%
\special{pn 13}%
\special{sh 1}%
\special{ar 3800 2500 10 10 0  6.28318530717959E+0000}%
\special{sh 1}%
\special{ar 3800 2500 10 10 0  6.28318530717959E+0000}%
%
\special{pn 13}%
\special{sh 1}%
\special{ar 4350 1950 10 10 0  6.28318530717959E+0000}%
\special{sh 1}%
\special{ar 4350 1940 10 10 0  6.28318530717959E+0000}%
%
\special{pn 13}%
\special{sh 1}%
\special{ar 3500 2800 10 10 0  6.28318530717959E+0000}%
\special{sh 1}%
\special{ar 3500 2800 10 10 0  6.28318530717959E+0000}%
%
\special{pn 13}%
\special{sh 1}%
\special{ar 1800 2800 10 10 0  6.28318530717959E+0000}%
\special{sh 1}%
\special{ar 1800 2800 10 10 0  6.28318530717959E+0000}%
%
\special{pn 13}%
\special{sh 1}%
\special{ar 3800 2800 10 10 0  6.28318530717959E+0000}%
\special{sh 1}%
\special{ar 3800 2800 10 10 0  6.28318530717959E+0000}%
%
\special{pn 8}%
\special{pa 3830 2930}%
\special{pa 3864 2930}%
\special{pa 3896 2932}%
\special{pa 3926 2938}%
\special{pa 3956 2946}%
\special{pa 3986 2960}%
\special{pa 4014 2976}%
\special{pa 4040 2994}%
\special{pa 4050 3000}%
\special{sp}%
\put(39.0000,-32.0000){\makebox(0,0)[lb]{$y=t_1x-\alpha_1=0$}}%
\put(54.0000,-18.0000){\makebox(0,0)[lb]{$y=t_3 x-\alpha_3 $}}%
\put(48.0000,-14.0000){\makebox(0,0)[rb]{$y=t_2 x-\alpha_2$}}%
\end{picture}%
 \caption{Three lines are the lagrangian submanifolds. 
We fix $\alpha_1=0$ by the translational invariance. 
For $t_1=0$, $t_2=\frac{p}{q}$ and $t_3=\frac{P}{Q}$, 
the area of the triangle surrounded by the three line 
is equal to the value $\triangle_0$ of 
$\mu=\nu=0$ and $\epsilon_i=\alpha_i$. 
}
 \label{fig:tri2}
\end{figure}
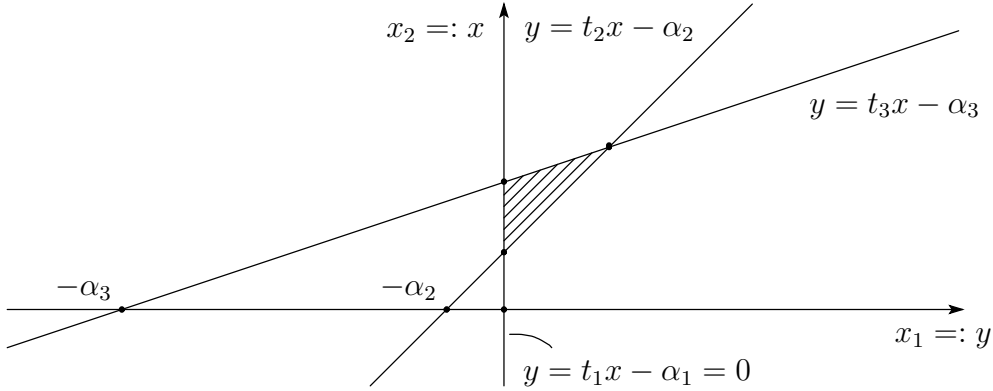
Note that $\triangle_m$ for each $m$, $\mu$, $\nu$ and $\epsilon_i$ 
for $i=1,2,3$ is just the area of a triangle surrounded by three 
geodesic cycles on the covering space. 
For instance if $\epsilon_i=\alpha_i$, $\alpha_i\in\R$ for $i=1, 2, 3$ 
and $m=0$, then $\triangle_0$ coincides with the area of the triangle 
surrounded by three lines $y=-\alpha_1$, $y=\frac{p}{q}x-\alpha_2$ 
and $y=\frac{P}{Q}x-\alpha_3$ 
where $x:=x_2, y:=x_1$ (see Fig.\ref{fig:tri2}). 
Next, for simplicity we set $\epsilon_i=0$ and $\mu=\nu=0$. 
In this case $\triangle_0=0$ since the corresponding three lines intersect 
at the origin. $\triangle_m$ is then the area surrounded by 
$y=0$, $y=\frac{p}{q}x$ and 
$y-pm=\frac{P}{Q}(x-qm)$. The vertex $(x,y)=(0,0)$ corresponds to 
$\epsilon=\mu=0$. The vertex $(x, y)=(qm,pm)$ corresponds to 
$\epsilon'=\nu=0$. While, the rest vertex $(x, y)=(0,\frac{p'm}{P})$ 
means that the vertex just corresponds to $\rho=p'm$ up to $P$ 
(see Eq.(\ref{strconst})). 
Fig.\ref{fig:tri} shows such a situation in the case 
$(p,q)=(1,1)$ and $(P, Q)=(3, 1)$ ($(p',q')=(2, 1)$). 
Note that 
the vertex $(x, y)=(0,\frac{p'm}{P})$ denotes the same point on the 
two-torus when $m$ is replaced to $m+P$. In this way, on two-tori 
there are infinitely many triangles whose vertices are common. 
The structure constant is then 
given by summing up all the triangles. 

Generally (but in the case of $t_1=\alpha_1=0$), 
$e_\mu$, $e_\nu$ and $e_\rho$ correspond to 
points 
$(x, y)=(\frac{q}{p}(\mu+\alpha_1), 0)$,  
$(x, y)=(\frac{q'q\nu+qQ(\alpha_2-\alpha_3)}{p'},
\frac{q'p\nu+qP\alpha_2-pQ\alpha_3}{p'})$ and 
$(x, y)=(\frac{Q}{P}(\rho+\alpha_3), 0)$, respectively, 
on the two-torus. 
Since the slope of the three lines $t_i$ are given, 
the triangle is obtained when the two points $e_\mu$ and $e_\nu$ are 
determined. The condition that the rest third vertex coincides with 
$e_\rho$ is then the $\delta_{R(p'm+Q\mu+q'\nu)}^\rho$ in 
Eq.(\ref{strconst2}). 
One can see that the place of these lines depend not only 
on $\alpha_i$ but on $\mu$ and $\nu$. 
For instance for the holomorphic vector which belongs to 
the $\cA_\theta$-$\cA_{\theta'}$ bimodule, 
the line of $i=2$ is described by 
$y=\frac{p}{q}(x-\frac{q\mu}{p})-\alpha_2$. 
This agrees with the viewpoints of the Heisenberg module in the previous 
subsection. 
In fact, we have identified $f(z,\mu)$ as the function on the line 
$y=\frac{p}{q}(x-\frac{q\mu}{p})$, which just coincides with the situation 
here
\footnote{In fact, in the previous subsection 
one could use other identifications 
which differ by the translation $\alpha_i$. }.  
This fact holds in a similar way 
for any holomorphic vector defined between any two D1-brane, 
since the arguments can reduce to the above ones by a $SL(2,\Z)$ 
transformation.

Let us denote $\epsilon_i=\alpha_i+\frac{\beta_i}{\tau}$ where 
$\alpha_i, \beta_i\in\R$. As explained above, 
$\alpha_i$ expresses the place of the D1-brane $i$. 
Then what is $\beta_i$? 
By comparing the holomorphic vector in Eq.(\ref{holvect}) and 
Eq.(\ref{holvect2}), we can see that it corresponds to $c_2$. 
(We can also see that $\alpha_i$ corresponds to $c_1$.) 
Furthermore it means that we consider the holomorphic vector with 
$\nabla_2=\fpartial{z}+2\pi i c_2$. This is nothing but the 
holonomy on the D1-brane parameterized by $z$. 
More precisely, $\beta_i$ corresponds to the holonomy on the 
corresponding D1-brane for each $i$. 
For instance, 
for the holomorphic vector which belongs to 
the $\cA_{\theta'}$-$\cA_{\theta''}$ module, 
the connection $\nabla_2$ is 
$\nabla_2=\fpartial{z}-2\pi i (q-p\theta)
\l(\frac{Q}{Q-P\theta}\beta_3-\frac{q}{q-p\theta}\beta_2\r)$. 
Here the constant term proportional to $\beta_3$ is regarded as 
an element in the center of $\cA_{\theta''}$, whereas 
that proportional to $\beta_2$ is realized 
as a constant in $\cA_{\theta'}$. 
The scaling $\frac{Q}{Q-P\theta}$ and $\frac{q}{q-p\theta}$ in front of 
$\beta_i$s are also consistent with the identification of the coordinate 
$z$ in Fig.\ref{fig:period}. 
Then, the term linear for $\beta_i$s in $\triangle_m$ 
coincides with the integral of the flat connection 
along the boundary of the triangle, \ie the three D1-branes. 
Note that the triangle is an open string disk, so 
the integral is equal to that of flat gauge field along 
the boundary of the disk. 
These contributions of $\alpha_i$ and $\beta_i$ to the structure constant 
$c_{\mu\nu}^\rho$ are just the definition of the $A_\infty$-category 
in \cite{F,mirror}. 
The terms of $\beta_i$ square in $\triangle_m$ can be absorbed into 
the definition of the expression of the holomorphic vectors, 
and one can see the correspondence of the arguments here with those 
in \cite{PZ}.

We have taken the tensor product between the `holomorphic' 
$\cA_\theta$-$\cA_{\theta'}$ bimodule and 
$\cA_{\theta'}$-$\cA_{\theta''}$ bimodule and obtained the 
$\cA_{\theta}$-$\cA_{\theta''}$ bimodule. 
{}From the viewpoints of the $A_\infty$-category in \cite{PZ}, 
the tensor product corresponds to the 
product $m_2$ between the morphisms in the following situation; 
both morphisms are 
the homomorphisms between flat {\it line bundles} 
on lagrangian submanifolds, 
and one of these three lagrangian submanifolds is fixed 
to be $y=0$ (or $y=-\alpha_1$). 
In order to get the category of the holomorphic vectors which 
corresponds to the $A_\infty$-category of \cite{PZ}, 
it is sufficient to consider the tensor product in 
a more general case, the tensor product between 
$\Mat_{n_1}(\cA_{\theta'})$-$\Mat_{n_2}(\cA_{\theta''})$ bimodule and 
$\Mat_{n_2}(\cA_{\theta''})$-$\Mat_{n_3}(\cA_{\theta'''})$ bimodule. 
Here $\Mat_{n}(\cA_\theta)$ denotes $n\times n$ matrix with $\cA_\theta$ 
entry, which is the endomorphism algebra of 
the rank $n$ free module over $\cA_\theta$. 
First, one can easily define the tensor product 
between an $\cA_{\theta'}$-$\cA_{\theta''}$ module 
and an $\cA_{\theta''}$-$\cA_{\theta'''}$ module over $\cA_\theta$ 
by regarding it as the tensor product over the noncommutative torus 
$\cA_{\theta'}$ and using Eq.(\ref{tensorprod}). 
The structure constant is then determined by summing up the area 
of the triangles as in Fig.\ref{fig:tri2} where $t_1\in\mathbb{Q}$. 
Next, to extend it further for 
$\Mat_{n_i}(\cA_{\theta'})$-$\Mat_{n_{i+1}}(\cA_{\theta''})$ bimodules, 
essentially one may multiply $V_i\otimes V_{i+1}^*$ to 
holomorphic vectors in the $\cA_{\theta'}$-$\cA_{\theta''}$ bimodule, 
where $V_i=\C^{n_i}$. 
However, the space of morphisms of the $A_\infty$-category in \cite{PZ} 
is larger. 
By comparing our arguments with those in \cite{PZ}, 
one can see how the space of the holomorphic vectors should be extended 
to recover any morphism of the $A_\infty$-category in \cite{PZ}. 
The connection $\nabla_2$ should be extended by replacing 
$\beta_i$ to $\beta_i+N_i$ where $N_i\in \mathrm{End}V_i$ is a 
constant indecomposable nilpotent matrix.  
We keep $\nabla_1$ unchanged. Therefore the curvature 
$[\nabla_1,\nabla_2]=\frac{2\pi i p'}{q'-p'\theta}$ is preserved 
through this modification. 
Then we may consider, as the extended holomorphic vectors, 
the solutions of the equation 
$(-\tau\nabla_1+\nabla_2)f=0$ where $f$ belongs to 
the $\Mat_{n_i}(\cA_{\theta'})$-$\Mat_{n_{i+1}}(\cA_{\theta''})$ bimodule.

In \cite{PZ}, the explicit form of $m_2$ is given in the case $p=p'=1$ 
so in this case one can immediately confirm 
that Eq.(\ref{strconst}) or Eq.(\ref{tri}) 
actually coincides with the result in \cite{PZ}. 
The associativity of the tensor product can be checked directly, 
but it is also realized geometrically as the way of separating 
each quadrangle into two triangles\cite{F,Phigher}.

To summarize, the $A_\infty$-category is defined by using 
lagrangian submanifolds, that is, the D1-branes. 
Similarly, the Heisenberg modules are also defined by the D1-branes. 
In this picture, $m_2$ in the $A_\infty$-structure is identified with 
the tensor product between bimodules. 
Namely, the open string interactions on noncommutative two-tori 
are given by counting holomorphic disks \ie disk instantons on two-tori.

\vspace*{0.2cm}

$\bullet$ {\it Additional comments}

Indeed we have seen that the category of holomorphic vectors 
corresponds to 
the $A_\infty$-category for general $\theta$. 
However note that the expression of the holomorphic vectors 
depends on $\theta$. 
As seen from the expression in Eq.(\ref{holvect}), 
holomorphic vectors exist only if $\frac{p}{q-p\theta}> 0$. 
Morphisms of the $A_\infty$-category are then in one-to-one correspondence 
with the holomorphic vectors only if $\theta=0$. 
Thus the category of the holomorphic vectors depends on $\theta$. 
In this sense, we can say that 
the category of the holomorphic vectors gives a noncommutative 
extension of the $A_\infty$-category 
(when higher products $m_i$ for $i\ge 3$ are also constructed). 
Moreover, the morphisms of the $A_\infty$-category correspond to theta 
functions which are the morphisms of the (derived) category of 
coherent sheaves. 
Therefore these holomorphic vectors can be regarded as 
certain noncommutative theta functions. 
In \cite{Stheta} on higher dimensional tori 
the holomorphic vectors are discussed from the point of view of 
noncommutative analogue of theta functions. 
For other literatures about noncommutative extensions of theta functions, 
see \cite{Wtheta,Mtheta}. 

As stated previously, 
the $A_\infty$-category possesses higher product $m_i$ for $i\ge 3$
\cite{Phigher}. 
For lagrangian submanifolds on the covering space of the two-tori, 
the higher products $m_i$ are constructed by summing over the 
$i+1$-gons in a similar way as $m_2$.  
The corresponding higher products of 
the holomorphic vectors can also be constructed, 
since the graphical realization of the holomorphic vectors is given 
in this paper. 
The relations of the $A_\infty$-structure are also realized as 
ways of separating polygons into two polygons\cite{F,Phigher}. 
It is shown that, on the derived category of coherent sheaves side, 
the higher product of the $A_\infty$-structure correspond to the higher 
Massey products\cite{F,Phigher}. 
Therefore, the higher products of the holomorphic vectors may be 
regarded as noncommutative higher Massey products.

Finally, we mention about a symmetry that the category of 
the holomorphic vectors has. 
In the commutative case $\theta=0$, 
$\epsilon_i$s have a translational invariance. 
This is a property which the morphisms of the $A_\infty$-category have. 
For $\theta\ne 0$ the structure constant (\ref{strconst}) still 
has the translational invariance, but each holomorphic vector 
does not. This means, in other words, 
that the holomorphic vectors have a symmetry corresponding to 
the translational invariance of $\epsilon_i$. 
When $\theta=0$ this symmetry becomes degenerate.

 \section{Conclusions and Discussions}
\label{sec:CD}

In this paper, 
we related the physics of the D1-brane on two-tori to 
the Kronecker foliation. 
The product of the field corresponding to the open strings 
ending on the D1-brane was identified 
with the crossed product of the rotation algebras. 
The algebra of the fields was then identified with the noncommutative 
two-tori represented by the rotation algebra. 
We showed that, in a particular degenerate metric, 
the two-torus is `foliated'. 
That is, when one probes the 
physics on the two-torus by a D1-brane on a 1-cycle, the two-torus is 
`foliated' by the open strings ending on the D1-brane. 
Such a situation is described by the the Kronecker foliation, 
so the different choice of the 1-cycle on which the probe D1-brane lies 
are related by the Morita equivalence of the noncommutative two-tori. 
Thus the Morita equivalence of noncommutative two-tori was realized 
from the D1-brane physics geometrically. 
The degenerate metric is T-dual to the metric in the Seiberg-Witten limit. 
It is known that the open string theory in the Seiberg-Witten limit is 
realized from the viewpoints of the deformation quantization. 
Therefore, we can conclude that the two representations of 
noncommutative two-tori, that in the deformation quantization and that 
in the crossed product, are related by T-duality.   
Such viewpoints are then applied to the homological mirror symmetry. 
After the tensor product between two Heisenberg modules is constructed, 
the tensor product between two holomorphic vectors is calculated 
and it is identified with the $m_2$ in the $A_\infty$-structure defined 
in \cite{PZ}. 
This implies that the interaction of two `holomorphic' open strings 
between noncommutative D-branes on noncommutative two-torus 
can be obtained from disk instanton contributions on two-tori. 
This results can be regarded as a 
`noncommutative homological mirror symmetry' on two-tori.

Though we discussed the arguments above on the special degenerate metric, 
one might generalize it in the case of any nondegenerate metric 
by introducing additional moduli $\Phi$\cite{PS} as in \cite{I}. 
For higher dimensional tori, 
the noncommutative expression of the open string physics is 
essentially obtained by 
quantizing open strings in general boundary conditions\cite{K}. 
However, it is expected that the homological mirror symmetry 
cannot understood by straightforward extensions of the two-dimensional 
case\cite{F,KaOr}. 
Therefore, even if the tensor product of projective modules is constructed, 
it is unclear that the category of the holomorphic vectors is 
related completely to the homological mirror symmetry. 
Conversely, the construction of the tensor product in higher dimensional 
tori might give an insight for 
the homological mirror symmetry for higher dimensional tori. 

Different connections between noncommutative tori and the homological 
mirror symmetry are found in \cite{FNC,So,SV,BS,KaOr} for example. 
In \cite{FNC} lagrangian foliations are considered 
as the objects of the $A_\infty$-category. 
In \cite{So,SV} in the degenerate limit of the complex structure 
its noncommutative compactification is discussed. 

Formally, if one considers the direct sum of all objects of an 
$A_\infty$-category, 
the morphisms are unified as the endomorphism 
algebra which acts on it. 
In such a way, formally an $A_\infty$-category can be regarded 
as an $A_\infty$-algebra of the endomorphism algebra
(for the definition of $A_\infty$-algebras and a deep relation of them to 
the field theory of open string, see \cite{K2}). 
Alternatively, by definition, any projective module over $\cA_\theta$ 
is obtained by multiplying an appropriate idemponent on free 
module $(\cA_\theta)^N$ for sufficiently large $N$. 
Bimodules are also obtained by multiplying appropriate idemponents 
on $\Mat_N(\cA_\theta)$ from left and right. 
This implies that the $A_\infty$-category is naturally embedded into 
the $A_\infty$-algebra of $\Mat_N(\cA_\theta)$. 
Thus, physically, 
the results of this paper might be relevant to nonperturbative 
effects of large $N$ field theory over $\cA_\theta$ 
through tachyon condensation process.

\begin{center}
\noindent{\large \textbf{Acknowledgments}}
\end{center}

I am very grateful to A.~Kato for helpful discussions and advice. 
I would also like to thank Y.~Terashima for valuable discussions.
The author is supported by JSPS Research Fellowships for Young
Scientists.

\end{document}